\documentclass[a4paper,11pt]{article}
\usepackage{jheppub} 
\usepackage{lineno}
\usepackage{dsfont}
\usepackage{calc}
\usepackage{tikz}
\usetikzlibrary{decorations.pathmorphing}

\allowdisplaybreaks

\newcommand{\normal}[1]{{\hspace{-0.1ex}:\hspace{-0.6ex} #1 \hspace{-0.6ex}:\hspace{0.2ex}}}
\newcommand{\s}{\hspace{0.2ex}}
\newcommand{\barred}[1]{\hspace{0.2ex}\overline{\hspace{-0.2ex} #1}}

\title{\boldmath AdS$_3$ String Stars at Pure NSNS Flux}

\author{Nicholas Agia and Daniel L.~Jafferis}
\affiliation{Center for the Fundamental Laws of Nature, Harvard University,
Cambridge, MA, USA}

\emailAdd{nagia@g.harvard.edu}
\emailAdd{jafferis@g.harvard.edu}

\abstract{We study Horowitz-Polchinski string stars in pure AdS$_3$ near the Hagedorn temperature using the technique of worldsheet conformal perturbation theory. Since the worldsheet CFT for pure AdS$_3$ is known exactly, our methodology provides a systematic way to construct Horowitz-Polchinski backgrounds to all orders in $\alpha'$. We explicitly construct the leading string star equations in a double expansion in temperature and WZW level $k$ which we then solve numerically.}

\begin{document}

\maketitle
\flushbottom

\section{Introduction}\label{intro}

The study of black holes is a rich and vast cornerstone of modern physics, providing not only connections between classical general relativity and the search for a quantum theory of gravity but also insights which have actual astrophysical implications. In string theory, the celebrated correspondence principle \cite{Horowitz96} proposes a continuous family of backgrounds which interpolates between black holes on one end and highly-excited thermal string configurations, or ``string stars'' as they are sometimes called, on the other end, with the correspondence point where one type of background turns into the other is roughly where the Schwarzschild radius in both cases approaches the string scale. In \cite{Horowitz97}, Horowitz and Polchinski derived the flat space string star background equations by obtaining the effective target space equations of motion for a thermal winding condensate and a gravitational backreaction on top of the thermal string background. This procedure requires being close to the Hagedorn temperature; as one approaches the Hagedorn temperature, one of the thermal winding string states becomes arbitrarily light, and it is this state which condenses and backreacts the geometry to form the string star \cite{Atick88, Kutasov05, Chen21, Balthazar221, Balthazar222}. String stars have been studied in other contexts over the last several decades, and they represent an important ingredient in string theory thermodynamics and understanding stringy properties of small black holes.

In asymptotically flat space, string stars require $d > 2$ noncompact spatial dimensions. Indeed, one should not expect a flat space correspondence principle in three spacetime dimensions, as there also do not exist asymptotically flat black holes in this number of dimensions. However, for asymptotically AdS spacetimes, there do exist three-dimensional black hole backgrounds, namely the BTZ black holes \cite{Banados92}. Thus, it is interesting to study string stars in $\mathrm{AdS}_3$ \cite{Lin07, Rangamani07}. The correspondence principle and associated Horowitz-Polchinski backgrounds have also been discussed in higher-dimensional $\mathrm{AdS}_{d+1}$ for $d > 2$ \cite{Urbach22}. In $\mathrm{AdS}_3$, the correspondence principle becomes slightly more subtle due to the fact that Euclidean thermal $\mathrm{AdS}_3$ and Euclidean BTZ are simply related by a modular transformation of the boundary torus, so the black hole side of the $\mathrm{AdS}_3$ correspondence principle should consist of a BTZ-like background that also has a winding condensate present. String stars in $\mathrm{AdS}_3$ with some amount of RR flux present were also studied recently in \cite{Urbach23} from the target space effective action approach.

Another reason why the $\mathrm{AdS_3}$ case deserves special attention is that the worldsheet CFT describing strings propagating on $\mathrm{AdS_3}$ with pure NSNS flux is known exactly, being given by the $\mathrm{SL}(2,\mathds{R})_k$ WZW model \cite{Maldacena001,Maldacena002,Maldacena01}, where the WZW level $k$ sets the AdS scale in terms of the string scale. To take advantage of the exact description, in this paper we study $\mathrm{AdS_3}$ string stars at pure NSNS flux from the worldsheet perspective. In particular, we show how the Horowitz-Polchinski string star solutions are obtained by starting with the Euclidean thermal $\mathrm{AdS_3}$ worldsheet CFT near the Hagedorn temperature and then deforming by time-winding and gravitational string vertex operators, solving for the desired deformation profiles via worldsheet conformal perturbation theory. This procedure holds to all orders in $\alpha'$, as long as one systematically includes the contributions from the massive string states when they become non-negligible. 

The worldsheet CFT and target space effective action approaches are, of course, equivalent, with the worldsheet beta functions essentially corresponding to the equations of motion of the target space effective action. Nevertheless, some aspects of the Horowitz-Polchinski methodology are clearer from the worldsheet perspective. For example, it is the worldsheet holomorphy of the preserved time-translation symmetry when acting on zero-winding operators that constrains the massless backreaction to take the form in the paper below. In particular, this holomorphy fixes a nontrivial combination of the spacetime metric and Kalb-Ramond field which is not obvious from the spacetime effective field theory approach.

The outline for the paper is as follows. In Section \ref{flat space}, we review the Horowitz-Polchinski methodology in flat space but from the perspective of the string worldsheet, which is perhaps less familiar than the traditional target space effective action approach. The bulk of the paper is contained in Section \ref{AdS HP}, where we systematically derive the Horowitz-Polchinski string star equations in $\mathrm{AdS_3}$. After reviewing the duality between the $\mathrm{SL}(2,\mathds{C})_k/\mathrm{SU}(2)$ WZW model describing Euclidean $\mathrm{AdS_3}$ and the uplifted sine-Liouville theory, we use the latter description to derive the leading beta functions of the winding condensate plus backreaction deformation. We then take the flat space limit of the result and use the results of Section \ref{flat space} to fix unambiguously the normalization of the sine-Liouville correlators, after which we take the semiclassical limit for AdS-scale string stars which are large compared to the string scale. In Section \ref{numerics}, we numerically solve the string star equations for various large $k$ and near-Hagedorn temperatures. We furthermore transform these solutions to the target space position basis to make contact with the form of the string star profiles one would expect to obtain from a target spacetime effective action approach. Finally, in Section \ref{stringy} we present the formulas needed to take the first several stringy corrections into account.

\section{Review of Flat Space Horowitz-Polchinski from the Worldsheet}\label{flat space}

Let us now review the asymptotically-flat case originally studied by Horowitz and Polchinski \cite{Horowitz96, Horowitz97} but from the perspective of the worldsheet CFT. The idea is to construct a valid CFT background near the Hagedorn temperature from a Euclidean winding condensate on top of thermal $S^1_{\beta}\times \mathds{R}^d$. We shall work in bosonic string theory for simplicity, and we shall not explicitly write the internal part of the matter CFT. Let $X^0$ be the Euclidean time compact boson at radius $R = \frac{\beta}{2\pi}$, and let $X^i$ be the spatial free bosons on $\mathds{R}^d$, for which the relevant part of the holomorphic stress-energy tensor is
\begin{equation}
T(z) = -\frac{1}{\alpha'}\normal{\partial X^0(z)\partial X^0(z)} - \frac{1}{\alpha'}\normal{\partial X_i(z)\partial X^i(z)},
\end{equation}
with the usual OPE $X^{\mu}(z,\barred{z})X^{\nu}(0) \sim -\frac{\alpha'}{2}\delta^{\mu\nu}\ln|z|^2$; the Euclidean time-translation current is $J^0 = \frac{i}{\sqrt{\alpha'}}\partial X^0$. Foliating the spatial $\mathds{R}^d$ as $\mathds{R}^+\times S^{d-1}$, we are interested only in operators invariant under the $\mathrm{SO}(d)$ of the angular $S^{d-1}$. The usual plane-wave basis of bosonic matter vertex exponential operators in the Euclidean theory with periodicity $X^0 \sim X^0 + \beta$ is given by 
\begin{align}
\mathcal{V}^{w,n}_{\vec{k}} & = \normal{e^{i[(\frac{2\pi n}{\beta} + \frac{w\beta}{2\pi\alpha'})X^0_{\text{L}} + (\frac{2\pi n}{\beta} - \frac{w\beta}{2\pi\alpha'})X^0_{\text{R}}]}}\normal{e^{i\vec{k}\cdot\vec{X}}}
\\ h & = \frac{\alpha'}{4}\left(\frac{2\pi n}{\beta} + \frac{w\beta}{2\pi\alpha'}\right)^2 + \frac{\alpha'\vec{k}^{\s 2}}{4}
\\ \widetilde{h} & = \frac{\alpha'}{4}\left(\frac{2\pi n}{\beta} - \frac{w\beta}{2\pi\alpha'}\right)^2 + \frac{\alpha'\vec{k}^{\s 2}}{4},
\end{align}
where $\vec{k} \in (\mathds{R}^{d})^*$.
As a description of the associated Lorentzian string state, interpreted in ordinary radial quantization for $w = 0$ (i.e.~scattering states) and in angular quantization for $w \neq 0$ (i.e.~stretched strings), the center-of-mass momentum relates to the energy and mass via the usual dispersion relation $\vec{k}^{\s 2} = E^2-m^2$. The thermal string spectrum associated to $w = 0$ coincides with the zero-temperature Lorentzian string spectrum (and so the tachyon divergence of bosonic string theory is trivially present but is irrelevant to us). The on-shell vertex operators with $w \neq 0$, on the other hand, have dimension
\begin{equation}
\Delta = \frac{w^2\beta^2}{8\pi^2\alpha'} - \frac{\alpha' m^2}{2} \stackrel{!}{=} 2.
\end{equation}
At very low temperatures ($\beta\rightarrow\infty$), the effective mass is strictly positive, and hence the string behaves causally. However, as the temperature increases, $\beta$ eventually decreases to a critical value $\beta_*$ beyond which the effective $m^2$ becomes negative, representing a tachyonic instability. This critical temperature occurs first for the winding $w = \pm 1$ states, and is hence determined by $\Delta_*^{w = \pm 1} = \frac{\beta_*^2}{8\pi^2\alpha'} \stackrel{!}{=} 2$. Thus, this critical temperature $T_* = \frac{1}{\beta_*}$ is given by
\begin{equation}
T_* = T_{\text{H}} = \frac{1}{4\pi\sqrt{\alpha'}},
\end{equation}
which is the familiar flat space Hagedorn temperature. This divergence is meaningful, and it appears in the superstring, for which flat space is a tachyon-free vacuum, in the same way (but is a factor of $\sqrt{2}$ larger); beyond this temperature the stretched string state which became massless must condense. The first state to condense in this way has the smallest available energy, here being zero, and thus the associated Euclidean operators have zero momentum and unit winding around the thermal circle.

The basic vertex operators which are neutral under the spatial $\mathrm{SO}(d)$ are
\begin{equation}
\mathcal{O}^{w,n}_p(z,\barred{z}) = \frac{(4\pi)^{(d+2)/4}\alpha'^{d/4}}{p^{(d-3)/2}}\sqrt{\frac{\Gamma(\frac{d}{2})}{\beta}}\int\frac{d^d k}{(2\pi)^d}\delta(\vec{k}^{\s 2} - p^2)\mathcal{V}^{w,n}_{\vec{k}}(z,\barred{z}),
\end{equation}
for $p \geqslant 0$. The coefficient here is chosen so that these operators are canonically normalized, namely their sphere two-point functions are
\begin{align}
\notag \hspace{30pt}&\hspace{-30pt}\left\langle\mathcal{O}^{w,n}_p(z_1,\barred{z}_1)\mathcal{O}^{w',n'}_{p'}(z_2,\barred{z}_2)\right\rangle
\\ \notag & = \frac{\omega_{d-1}\beta\alpha'^{d/2}\delta_{n+n',0}\delta_{w+w',0}}{(2\pi)^d z_{12}^{2h}\barred{z}_{12}^{2\widetilde{h}}}\left[\frac{(4\pi)^{(d+2)/2}\Gamma(\frac{d}{2})}{(pp')^{(d-3)/2}\beta}\right]\int_0^{\infty} \hspace{-2pt}d|\vec{k}| \ |\vec{k}|^{d-1}\delta(|\vec{k}|^2 \hspace{-2pt}-\hspace{-2pt} p^2)\delta(|\vec{k}|^2 \hspace{-2pt}-\hspace{-2pt} p'^2)
\\ & = \frac{2\pi\alpha'^{d/2}\delta_{n+n',0}\delta_{w+w',0}\delta(p - p')}{z_{12}^{2h}\barred{z}_{12}^{2\widetilde{h}}},
\end{align}
where
\begin{equation}
\omega_{d-1} = \frac{2\pi^{d/2}}{\Gamma(\frac{d}{2})}
\end{equation}
is the volume of a unit $(d-1)$-sphere. Note that we have chosen the normalization of the free boson correlators to be set by the volume of the target space $S^1_{\beta}\times\mathds{R}^d$, namely $\langle\mathds{1}\rangle = (2\pi)^d\beta\delta^d(\vec{k}=0)$; the operators $\mathcal{O}^{w,n}_p$ have then been defined to carry no engineering dimension, and the correlator itself always carries the units of $(d+1)$-dimensional volume. The vertex operators of interest to us here are
\begin{align}
\mathcal{O}^{\pm}_p(z,\barred{z}) & \equiv \mathcal{O}^{\pm 1,0}_p(z,\barred{z})
\\ \mathcal{O}_p(z,\barred{z}) & \equiv 2\s\s\normal{J^0(z)\widetilde{J}^{\s 0}(\barred{z})\mathcal{O}^{0,0}_p(z,\barred{z})}
\end{align}
corresponding to the lightest states of unit time-winding and to the time-time component of the graviton, respectively; their dimensions are
\begin{align}
\Delta^{\pm}_p & = 2 + \frac{\alpha'p^2}{2} + \frac{2(\beta^2-\beta_{\text{H}}^2)}{\beta_{\text{H}}^2}
\\* \Delta_p & = 2 + \frac{\alpha'p^2}{2}.
\end{align}
In the flat space background, these vertex operators are off-shell for any nonzero momentum $p$.

\begin{figure}
\centering
\begin{tikzpicture}
\draw[gray,-stealth] (0,0) -- (0,5) node[above]{$T$};
\draw[gray] (0.1,4) -- (-0.1,4) node[left]{$T_{\text{H}}$};
\draw[gray,loosely dashed] (0,4) -- (9,4);
\node[scale=0.8] (1) at (1.2,2.4) {thermal};
\node[scale=0.8] (2) at (1.2,2) {string};
\node[scale=0.8] (3) at (1.2,1.6) {theory};
\draw[line width = 1pt] (1.8,0) -- (1.8,4);
\draw[line width = 1pt] (2,4) -- node[midway,above,sloped,scale=0.8,yshift=-0.1cm]{string star} (3.5,2.5);
\draw[dashed,line width=1pt] (2,4) -- (1,5);
\draw[line width = 1pt] (7.5,0.3) -- node[midway,above,sloped,scale=0.8]{black hole} (9,0);
\draw[dashed] (3.5,2.5) to[out=-45,in={180-atan(1/5)}] (7.5,0.3);
\draw[red,decorate,decoration={zigzag,amplitude=0.05cm,segment length = 0.12cm}] (1.8,3) -- (3,3);
\draw[red!50,-stealth] (2.9,2) -- (2.5,2.8);
\node[scale=0.8,red!50] (4) at (3,1.8) {off-shell};
\node[scale=0.8,red!50] (5) at (3,1.5) {deformation};
\end{tikzpicture}
\caption{A cartoon of the Horowitz-Polchinski methodology. Starting from the pure thermal string theory background near the Hagedorn temperature, one performs a deformation by slightly off-shell vertex operators led by the time-winding and massless states. One then solves the beta function equations to find a new (on-shell) string background finitely separated from the thermal starting point, which is the string star solution. The correspondence principle states that the string stars lie on one end of a line of string saddles connected to the black hole backgrounds.}
\label{string star}
\end{figure}
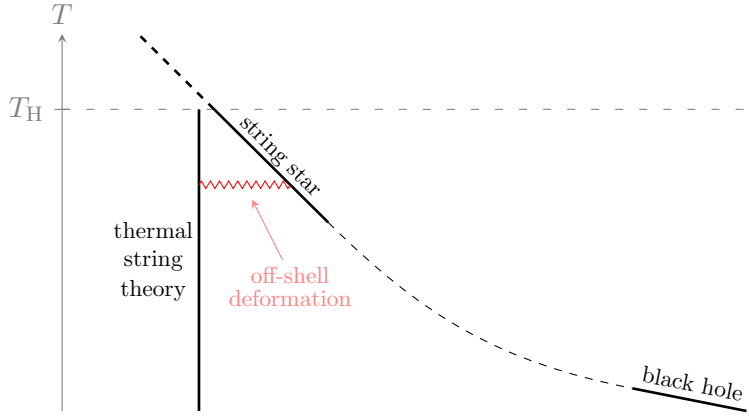

The logic of Horowitz and Polchinski is to find a new Euclidean string background by deforming the thermal string theory on $S^1_{\beta}\times\mathds{R}^d$ by a winding condensate and a massless backreaction near the Hagedorn temperature. The new background can be obtained from conformal perturbation theory by demanding that the deformed theory define a CFT. If one deforms a 2d CFT via
\begin{equation}
S = S_0 + \sum_i \lambda^i\int d^2 z \ \mathcal{O}_i(z,\barred{z})
\end{equation}
where $\lambda^i$ is a small coupling constant for each deforming operator $\mathcal{O}_i$ of dimension $\Delta_i$, then it has been known for a long time \cite{Cardy89, Polchinski982} that the leading-order beta function for $\lambda^i$ is
\begin{equation}\label{beta function compact}
\beta^i = (2-\Delta_i)\lambda^i + 2\pi\sum_{j,k}C_{jk}^{\phantom{jk}i}\lambda^j\lambda^k + \dotsc,
\end{equation}
where $C_{ij}^{\phantom{ij}k}$ are the OPE coefficients among the deformation operators. In the case of a conformal manifold, one starts with initially marginal operators ($\Delta_i = 2$) and then demands that the beta functions vanish order-by-order so that the deformation is exactly marginal. In the Horowitz-Polchinski scenario, on the other hand, one starts with slightly irrelevant operators ($\Delta_i > 2$) and solves $\beta^i = 0$ for finite deformation couplings. Specifically, one starts with the flat thermal background and considers the deformation
\begin{align}
S & = S_{S^1_{\beta}\times\mathds{R}^d} + \int d^2 z\int_0^{\infty}\frac{dp}{2\pi}\left[f^-(p)\mathcal{O}^+_p(z,\barred{z}) + f^+(p)\mathcal{O}^-_p(z,\barred{z}) + f(p)\mathcal{O}_p(z,\barred{z})\right] + \dotsc,
\end{align}
where $f^+(p) = f^-(p)^*$ and $f(p)$ are coupling functions to be determined, and the ``$+\dotsc$'' denotes deformations by more massive string states that may have to be turned on at higher order in $\alpha'$. The new background found by solving the beta function equations $\beta^{\pm}(p) = \beta(p) = 0$ then describes an on-shell background in which there is a thermal condensate of stretched strings together with a gravitational backreaction, called a ``string star''; see Figure \ref{string star}.
Horowitz and Polchinski argued \cite{Horowitz96} that there exists a continuous family of string saddles, parametrized by the string coupling, which interpolates between a macroscopic black hole and a small string star which they described thermodynamically via a random walk of a highly excited string; this proposal is known as the ``correspondence principle'', and is indicated by the dashed curve in Figure \ref{string star}. Note that it is not \emph{a priori} obvious if the string star branch should meet the thermal branch at the Hagedorn temperature in Figure \ref{string star}. Indeed, we shall find evidence that, at least for $\mathrm{AdS_3}$, the string star is finitely separated from the thermal background exactly at the Hagedorn temperature, which can even be extended to a valid background \emph{above} the Hagedorn temperature.

The beta function \eqref{beta function compact} was written for a discrete spectrum, but it immediately generalizes to the continuous case. For the normalization of the operators employed here, the beta functionals for the coupling functions $f^{\pm}(p)$ and $f(p)$ are
\begin{align}
\beta^{\pm}(p) & = \left(2-\Delta_p^{\pm}\right)f^{\pm}(p) + \frac{4\pi}{\alpha'^{d/2}}\int_0^{\infty}\frac{dp'}{2\pi}\int_0^{\infty}\frac{dp''}{2\pi}\mathcal{C}_{p,p',p''}f^{\pm}(p')f(p'') + \dotsc
\\* \beta(p) & = \left(2-\Delta_p\right)f(p) + \frac{4\pi}{\alpha'^{d/2}}\int_0^{\infty}\frac{dp'}{2\pi}\int_0^{\infty}\frac{dp''}{2\pi}\mathcal{C}_{p,p',p''}f^+(p')f^-(p'') + \dotsc,
\end{align}
where $\mathcal{C}_{p_1,p_2,p_3}$ is the three-point coefficient appearing in
\begin{equation}
\left\langle\mathcal{O}^+_{p_1}(z_1,\barred{z}_1)\mathcal{O}^-_{p_2}(z_2,\barred{z}_2)\mathcal{O}_{p_3}(z_3,\barred{z}_3)\right\rangle = \frac{\mathcal{C}_{p_1,p_2,p_3}}{|z_{12}|^{\Delta_{p_1}^++\Delta_{p_2}^- - \Delta_{p_3}}|z_{23}|^{-\Delta_{p_1}^++\Delta_{p_2}^- + \Delta_{p_3}}|z_{31}|^{\Delta_{p_1}^+-\Delta_{p_2}^- + \Delta_{p_3}}}.
\end{equation}
Using the OPEs
\begin{align}
J^0(z)\mathcal{O}^{\pm}_{p}(0) & \sim \pm\frac{\beta}{4\pi\sqrt{\alpha'} z}\mathcal{O}^{\pm}_{p}(0)
\\ \widetilde{J}^{\s 0}(\barred{z})\mathcal{O}^{\pm}_{p}(0) & \sim \mp\frac{\beta}{4\pi\sqrt{\alpha'} \barred{z}}\mathcal{O}^{\pm}_{p}(0),
\end{align}
it is straightforward to compute
\begin{multline}
\mathcal{C}_{p_1,p_2,p_3} = -\frac{2(4\pi)^{(3d-2)/4}[\Gamma(\frac{d}{2})\beta]^{3/2}\alpha'^{(3d-4)/4}}{(p_1 p_2 p_3)^{(d-3)/2}}
\\ \times \int \frac{d^d k_1}{(2\pi)^d}\int \frac{d^d k_2}{(2\pi)^d}\int \frac{d^d k_3}{(2\pi)^d}\delta(\vec{k}_1^{\s 2} - p_1^2)\delta(\vec{k}_2^{\s 2} - p_2^2)\delta(\vec{k}_3^{\s 2} - p_3^2)(2\pi)^d\delta^d(\vec{k}_1 + \vec{k}_2 + \vec{k}_3).
\end{multline}
Performing the integrals, this three-point coefficient is found to be
\begin{multline}\label{flat space 3-point coefficient}
\mathcal{C}_{p_1,p_2,p_3} = -\frac{[\Gamma(\frac{d}{2})\frac{\beta}{2\pi}]^{3/2}\alpha'^{(3d-4)/4}}{(d-2)!\pi^{d/4}}\theta\big(\Delta(p_1,p_2,p_3)\big)
\\ \times\left(\frac{(p_1\hspace{-2pt}+\hspace{-2pt}p_2\hspace{-2pt}+\hspace{-2pt}p_3)(-p_1\hspace{-2pt}+\hspace{-2pt}p_2\hspace{-2pt}+\hspace{-2pt}p_3)(p_1\hspace{-2pt}-\hspace{-2pt}p_2\hspace{-2pt}+\hspace{-2pt}p_3)(p_1\hspace{-2pt}+\hspace{-2pt}p_2\hspace{-2pt}-\hspace{-2pt}p_3)}{2p_1 p_2 p_3}\right)^{(d-3)/2},
\end{multline}
where $\theta\big(\Delta(p_1,p_2,p_3)\big)$ is a Heaviside theta function which is unity when $p_1$, $p_2$ and $p_3$ satisfy the triangle inequality and vanishes otherwise.

In terms of the dimensionless coupling functions
\begin{equation}
g(q) \equiv \frac{1}{\sqrt{\alpha'}}f\left(\frac{q}{\sqrt{\alpha'}}\right)
\end{equation}
and the dimensionless inverse temperature parameter
\begin{equation}
\epsilon \equiv \frac{\beta-\beta_{\text{H}}}{\beta_{\text{H}}},
\end{equation}
the vanishing of the above beta functions results in the equations
\begin{align}\label{flat eq 1}
0 & \stackrel{!}{=} \left[q^2 + 4\epsilon(\epsilon+2)\right]g^{\pm}(q) - \frac{2}{\pi}\int_0^{\infty}dq'\int_0^{\infty}dq'' \ \hat{\mathcal{C}}_{q,q',q''}g^{\pm}(q')g(q'')
\\ \label{flat eq 2} 0 & \stackrel{!}{=} q^2 g(q) - \frac{2}{\pi}\int_0^{\infty}dq'\int_0^{\infty}dq'' \ \hat{\mathcal{C}}_{q,q',q''}g^+(q')g^-(q''),
\end{align}
where
\begin{multline}
\hat{\mathcal{C}}_{q_1,q_2,q_3} = -\frac{[2\Gamma(\frac{d}{2})(1+\epsilon)]^{3/2}}{(d-2)!\pi^{d/4}}\theta\big(\Delta(q_1,q_2,q_3)\big)
\\ \times \left(\frac{(q_1\hspace{-2pt}+\hspace{-2pt}q_2\hspace{-2pt}+\hspace{-2pt}q_3)(-q_1\hspace{-2pt}+\hspace{-2pt}q_2\hspace{-2pt}+\hspace{-2pt}q_3)(q_1\hspace{-2pt}-\hspace{-2pt}q_2\hspace{-2pt}+\hspace{-2pt}q_3)(q_1\hspace{-2pt}+\hspace{-2pt}q_2\hspace{-2pt}-\hspace{-2pt}q_3)}{2q_1 q_2 q_3}\right)^{(d-3)/2}.
\end{multline}
A string star corresponds to a normalizable solution to \eqref{flat eq 1} and \eqref{flat eq 2} because only a normalizable deformation does not change the asymptotics of the original background. In the original Horowitz-Polchinski work \cite{Horowitz97}, the new background near the Hagedorn temperature with a winding condensate was obtained from the target space effective field theory equations of motion. In particular, Horowitz and Polchinski obtained a nonlinear, but local, differential equation whose solution gives the winding condensate profile directly in position space. The equations \eqref{flat eq 1} and \eqref{flat eq 2} as derived from the worldsheet take the form of nonlinear integral equations, which generically are in no way local. Nevertheless, we shall now show that \eqref{flat eq 1} and \eqref{flat eq 2} can actually be transformed to local differential equations.

Recall that, in $d$ spatial dimensions, the complete basis of functions with total angular momentum $\ell$ and definite radial momentum $q$ has radial dependence given by
\begin{equation}
\varphi_{q,\ell}(r) = \sqrt{2\pi qr} \s\s J_{\ell+\frac{d}{2}-1}(qr),
\end{equation}
normalized according to
\begin{equation}
\int_0^{\infty}dr \ \varphi_{q,\ell}(r)\varphi_{q',\ell}(r) = 2\pi\delta(q-q').
\end{equation}
Thus, the ``position space'' Horowitz-Polchinski profiles $\tilde{g}(r)$ are related to the coupling functions $g(q)$ via
\begin{equation}
g(q) = \int_0^{\infty}dr\sqrt{2\pi qr}\s\s \tilde{g}(r)J_{\frac{d}{2}-1}(qr),
\end{equation}
which is a Fourier-Bessel transform with $\ell=0$. In substituting this transform into \eqref{flat eq 1} and \eqref{flat eq 2}, the requisite integrals can actually be performed in closed form, thanks to the identity\footnote{See, for instance, A.P.~Prudnikov, Yu.~A.~Brychkov and O.I.~Marichev, \emph{Integrals and Series}, \emph{Volume 2 -- Special Functions}, subsection 2.12.6, identity 1.}
\begin{multline}
\int_{|a-b|}^{a+b} dx \sqrt{x}\left(\frac{[(a+b)^2-x^2][x^2-(a-b)^2]}{2x}\right)^{(d-3)/2}J_{\frac{d}{2}-1}(cx)
\\ = \sqrt{\frac{\pi}{2}}\left(\frac{4ab}{c}\right)^{(d-2)/2}\Gamma\left(\frac{d-1}{2}\right)J_{\frac{d}{2}-1}(ac)J_{\frac{d}{2}-1}(bc).
\end{multline}
Specifically, we have
\begin{multline}
\int_0^{\infty}dq'\int_{|q'-q|}^{q'+q}dq'' \ g^+(q')g^-(q'')\left(\frac{(q\hspace{-2pt}+\hspace{-2pt}q'\hspace{-2pt}+\hspace{-2pt}q'')(-q\hspace{-2pt}+\hspace{-2pt}q'\hspace{-2pt}+\hspace{-2pt}q'')(q\hspace{-2pt}-\hspace{-2pt}q'\hspace{-2pt}+\hspace{-2pt}q'')(q\hspace{-2pt}+\hspace{-2pt}q'\hspace{-2pt}-\hspace{-2pt}q'')}{2qq'q''}\right)^{(d-3)/2}
\\ = 2^{d-2}\pi\Gamma\left(\frac{d-1}{2}\right)\int_0^{\infty}dr\frac{\tilde{g}^+(r)\tilde{g}^-(r)}{r^{(d-1)/2}}\sqrt{2\pi qr}J_{\frac{d}{2}-1}(qr).
\end{multline}
Then, since
\begin{equation}
q^2 g(q) = -\int_0^{\infty}dr \sqrt{2\pi qr}\left(\frac{d^2\tilde{g}(r)}{dr^2} - \frac{(d-1)(d-3)}{4r^2}\tilde{g}(r)\right) J_{\frac{d}{2}-1}(qr),
\end{equation}
the leading string star equations in the position basis, after multiplying through by $\frac{1}{r^{(d-1)/2}}$, become
\begin{align}\label{flat eq 3}
0 & \stackrel{!}{=} -\frac{1}{r^{d-1}}\frac{d}{dr}\left[r^{d-1}\frac{d}{dr}\left(\frac{\tilde{g}^{\pm}(r)}{r^{(d-1)/2}}\right)\right] + 4\epsilon(\epsilon\hspace{-2pt}+\hspace{-2pt}2)\frac{\tilde{g}^{\pm}(r)}{r^{(d-1)/2}} + \frac{4\sqrt{2\Gamma(\frac{d}{2})(1\hspace{-2pt}+\hspace{-2pt}\epsilon)^3}}{\pi^{(d-2)/4}}\frac{\tilde{g}^{\pm}(r)}{r^{(d-1)/2}}\frac{\tilde{g}(r)}{r^{(d-1)/2}}
\\ \label{flat eq 4} 0 & \stackrel{!}{=} -\frac{1}{r^{d-1}}\frac{d}{dr}\left[r^{d-1}\frac{d}{dr}\left(\frac{\tilde{g}(r)}{r^{(d-1)/2}}\right)\right] + \frac{4\sqrt{2\Gamma(\frac{d}{2})(1\hspace{-2pt}+\hspace{-2pt}\epsilon)^3}}{\pi^{(d-2)/4}}\frac{\tilde{g}^+(r)}{r^{(d-1)/2}}\frac{\tilde{g}^-(r)}{r^{(d-1)/2}},
\end{align}
demonstrating that \eqref{flat eq 1} and \eqref{flat eq 2} are indeed local equations in spacetime! Moreover, \eqref{flat eq 3} and \eqref{flat eq 4} are precisely the differential equations that Horowitz and Polchinski obtained in \cite{Horowitz97} (their equations (3.12) and (3.13)), where their functions are given (modulo normalization) in terms of ours by $\chi(r) = \frac{\tilde{g}^+(r)}{r^{(d-1)/2}}$ for the winding condensate profile and $h(r) = \frac{\tilde{g}(r)}{r^{(d-1)/2}}$ for the graviton backreaction profile.

\section{\boldmath AdS\texorpdfstring{$_3$}{3} HP Equation}\label{AdS HP}

We would now like to construct string star solutions in $\mathrm{AdS_3}$ using the Horowitz-Polchinski methodology. We again work in bosonic string theory since the tachyon plays no role here, and moreover the leading structure of the solutions will still hold in the superstring case. It is well known that the worldsheet CFT describing string theory on (Lorentzian) $\mathrm{AdS_3}$ is given by the $\mathrm{SL}(2,\mathds{R})_k$ WZW model \cite{Maldacena001,Maldacena002,Maldacena01}, where the level $k$ sets the AdS length $\ell_{\mathrm{AdS}}$ in terms of the string length $\ell_{\text{s}} = \sqrt{\alpha'}$ via
\begin{equation}
\frac{\ell_{\mathrm{AdS}}}{\ell_{\text{s}}} = \sqrt{k-2}.
\end{equation}
Similarly, the worldsheet theory describing Euclidean $\mathrm{AdS_3}$ is the $\mathrm{SL}(2,\mathds{C})_k/\mathrm{SU}(2)$ WZW model, and its orbifold $\mathds{Z}_{\beta}\backslash \mathrm{SL}(2,\mathds{C})_k/\mathrm{SU}(2)$ which compactifies the Euclidean time on a circle of periodicity $\beta$ describes thermal $\mathrm{AdS_3}$ at inverse temperature $\beta$. The Euclidean BTZ background is then related to the thermal $\mathrm{AdS_3}$ background by a modular transformation of the boundary torus, which in particular sends $\beta \mapsto \frac{4\pi^2}{\beta}$. Just as in flat space, the partition function for string theory on thermal $\mathrm{AdS_3}$ diverges at a critical temperature. From the explicit expression for the partition function on thermal $\mathrm{AdS_3}$ computed in \cite{Maldacena002}, the exact $\mathrm{AdS_3}$ Hagedorn inverse temperature is computed as\footnote{\emph{Nota Bene}: In many places in the literature, there is an overall factor of $k$ in the $\mathrm{AdS_3}$ metric that is not present in the Minkowski metric, so that the flat space coordinates are $\sqrt{k}$ times the $\mathrm{AdS_3}$ coordinates in the $k\rightarrow \infty$ limit. Here we take the physical metric, meaning the one whose coordinates do become the flat space coordinates as $k\rightarrow\infty$. Indeed, here we have $\beta_{\text{H}} \stackrel{k\rightarrow\infty}{\longrightarrow} 4\pi\sqrt{\alpha'}$, the familiar flat space Hagedorn inverse temperature, whereas elsewhere in the literature this limit would be written $\sqrt{k}\beta_{\text{H}} \stackrel{k\rightarrow\infty}{\longrightarrow} 4\pi\sqrt{\alpha'}$.}
\begin{equation}\label{AdS Hagedorn}
\beta_{\text{H}} = 4\pi\sqrt{\alpha'\left(1-\frac{1}{4(k-2)}\right)}.
\end{equation}
The Horowitz-Polchinski process is to deform the string theory on thermal $\mathrm{AdS_3}$ near the Hagedorn temperature by a time-winding condensate of the Euclidean string state which becomes massless exactly at $\beta_{\text{H}}$ and by a graviton backreaction; the string star corresponds to the finite deformation thereof which again defines a CFT. Performing this procedure via conformal perturbation theory on the worldsheet has the advantage of being $\alpha'$-exact, but the $\mathds{Z}_{\beta}\backslash \mathrm{SL}(2,\mathds{C})_k/\mathrm{SU}(2)$ orbifold CFT relevant for thermal $\mathrm{AdS_3}$ is far less completely understood than its $\mathrm{SL}(2,\mathds{C})_k/\mathrm{SU}(2)$ parent or $\mathrm{SL}(2,\mathds{R})_k$ cousin. For instance, we must be able to describe the relevant time-winding operator in the abstract CFT as well as be able to compute its three-point coefficients with its conjugate and a third operator in the original untwisted sector. We find it simpler to work instead with an asymptotically-free-field description that is dual to the $\mathrm{SL}(2,\mathds{C})_k/\mathrm{SU}(2)$ CFT in terms of which the orbifold to thermal $\mathrm{AdS_3}$ becomes trivial.

\subsection{Review of AdS\texorpdfstring{$_3$}{3} Duality}

The duality proposed in \cite{Jafferis21} is based on uplifting the FZZ duality, the latter originally between the $\mathrm{SL}(2,\mathds{R})_k/\mathrm{U}(1)$ WZW model and sine-Liouville theory \cite{FZZ}, effectively ``ungauging'' the $\mathrm{U}(1)$. This proposal was subsequently checked in more detail in \cite{Halder22}. Here, we shall review only those parts of the duality needed for our computations.

The uplifted dual sine-Liouville worldsheet action in flat coordinates is
\begin{equation}\label{sL action}
S_{\text{sL}} = \frac{1}{2\pi\alpha'}\int d^2 z \left[\partial\phi\barred{\partial}\phi + \beta\barred{\partial}\gamma + \tilde{\beta}\partial\tilde{\gamma} + 2\pi\alpha'\mu(\mathcal{W}_+ + \mathcal{W}_-)\right] - Q\chi\phi_0,
\end{equation}
where $\phi$ is a linear dilaton with background charge $-Q$,
\begin{equation}
Q \equiv \frac{1}{\sqrt{\alpha'(k-2)}},
\end{equation}
the (commuting) $\beta\gamma$ system has conformal weights $h_{\beta} = 1$ and $h_{\gamma} = 0$ and the operators $\mathcal{W}_{\pm}$ are suitable uplifts of the sine-Liouville operator which have the interpretation of having winding $\pm 1$ around the spatial $\mathrm{AdS_3}$ circle; the usual topological term involving the dilaton zero-mode $\phi_0$ is proportional to the Euler characteristic $\chi = 2 - 2g$ for a genus $g$ worldsheet. The fundamental free OPEs are
\begin{align}
\phi(z,\barred{z})\phi(0) & \sim -\frac{\alpha'}{2}\ln|z|^2
\\ \gamma(z)\beta(0) & \sim \frac{\alpha'}{z},
\end{align}
with $\gamma(z)\gamma(0)$ and $\beta(z)\beta(0)$ nonsingular. The holomorphic stress-energy tensor is
\begin{equation}
T = -\frac{1}{\alpha'}\normal{\partial\phi\partial\phi} - Q\partial^2\phi - \frac{1}{\alpha'}\normal{\beta\partial\gamma},
\end{equation}
whose central charge
\begin{equation}
c = 1 + 6\alpha'Q^2 + 2 = \frac{3k}{k-2}
\end{equation}
immediately agrees with that of the $\mathrm{SL}(2,\mathds{R})_k$ WZW model. The basic family of primaries are constructed as exponentials of $\phi$, $\gamma$ and $\beta$; the former two are ordinary exponentials of a weight-zero field. For the latter, an expression like $e^{im\int_C^z dz'\beta(z')}$ represents a defect operator defined by the line integral. The associated branch cut in any OPE or correlator is defined by the defect line, with the defect operator itself defined by the equality
\begin{equation}
e^{in\gamma(z)}e^{im\int_C^0 dz'\beta(z')} = z^{\alpha'nm}\normal{e^{in\gamma(z)}e^{im\int_C^0 dz'\beta(z')}}.
\end{equation}
One need not be too concerned about the defect nature of such an operator, as the exponentials that are relevant to us are in fact local operators, with the branch cut canceling between the holomorphic and antiholomorphic contributions. Then, the general holomorphic exponential operator is written as
\begin{equation}
\mathcal{V}_{\alpha,n,m}(z) \equiv \normal{e^{in\gamma(z)}e^{im\int_C^z dz'\beta(z')}}\normal{e^{2\alpha\phi(z)}},
\end{equation}
whose holomorphic weight (in the free theory) is
\begin{equation}
h_{\alpha,n,m} = -\alpha'\alpha(Q+\alpha) + \alpha'nm.
\end{equation}

We have yet to give a precise definition of the sine-Liouville operators $\mathcal{W}_{\pm}$ appearing in \eqref{sL action}. In order to do so, we must introduce the $\mathrm{SL}(2,\mathds{R})_k$ currents $J^a(z)$, which obey the usual Ka\v{c}-Moody algebra
\begin{equation}
J^a(z)J^b(0) \sim \frac{k\delta^{ab}}{z^2} + \frac{if^{ab}_{\phantom{ab}c}}{z}J^c(0),
\end{equation}
where $\delta^{ab}$ is the Killing metric on $\mathfrak{sl}(2,\mathds{R})$. In the spin basis, $a \in \{\pm,3\}$, the nonzero components of the Killing metric are $\delta^{33} = -\frac{1}{2}$, $\delta^{+-} = \delta^{-+} = 1$, and the Cartan generators are
\begin{align}\label{J3 hol}
J^3 &  = \sqrt{\frac{k}{\alpha'}}\left(\beta + \frac{1}{4}\partial\gamma\right)
\\ \label{J3 antihol} \widetilde{J}^{\s 3} & = \sqrt{\frac{k}{\alpha'}}\left(\tilde{\beta} + \frac{1}{4}\barred{\partial}\tilde{\gamma}\right).
\end{align}
We shall not need the other currents $J^{\pm}$ in what follows. In the NL$\sigma$M description of thermal $\mathrm{AdS_3}$, the field $\gamma$ is the target complex coordinate on the boundary torus whose real component runs along the thermal circle and whose imaginary component runs along the spatial circle. That is, $\gamma$ is torus-valued with the periodicities $\gamma \sim \gamma + 2\pi i\sqrt{\alpha'k}$ and $\gamma\sim \gamma+\beta$, where here $\beta$ is the inverse temperature\footnote{We apologize for the notational clash of having $\beta$ denote both a worldsheet field and a spacetime inverse temperature. It should be obvious from context which meaning is intended.}. Since $J^3$ appears to shift the complex coordinate $\gamma$, one might expect that the current which generates translations around the thermal circle should be $J^3+\widetilde{J}^{\s 3}$ and that the current which generates translations around the spatial circle should be $J^3 - \widetilde{J}^{\s 3}$. However, this na\"{i}ve guess is not correct. The (anti-)holomorphic currents in \eqref{J3 hol} and \eqref{J3 antihol} are indeed those which shift the \emph{cylinder}-valued complex coordinate $\gamma$ in the Euclidean $\mathrm{AdS_3}$ described by the $\mathrm{SL}(2,\mathds{C})_k/\mathrm{SU}(2)$ WZW model. However, we are working with a dual description of the orbifold $\mathds{Z}_{\beta}\backslash \mathrm{SL}(2,\mathds{C})_k/\mathrm{SU}(2)$ instead, and one may consider deforming the currents \eqref{J3 hol} and \eqref{J3 antihol} by total derivatives which act nontrivially only on the twisted sectors of the orbifold. In this way, it is not \emph{a priori} obvious what the correct boundary torus translation currents are. Since $\gamma + \tilde{\gamma}$ is the coordinate along the time direction, the potential ambiguity lies in total derivatives of the form $\partial(\gamma+\tilde{\gamma})$, so we may consider the family of operators
\begin{align}\label{J3 nonhol}
\mathcal{J}^3 = \beta + \frac{1}{4}\partial\gamma + \frac{\varepsilon}{4}\partial\left(\gamma+\tilde{\gamma}\right)
\\ \label{J3 nonantihol} \widetilde{\mathcal{J}}^{\s 3} = \tilde{\beta} + \frac{1}{4}\barred{\partial}\tilde{\gamma} + \frac{\varepsilon}{4}\barred{\partial}\left(\gamma+\tilde{\gamma}\right)
\end{align}
for some real parameter $\varepsilon$. To reiterate the distinction, $J^3$ and $\widetilde{J}^{\s 3}$ are the holomorphic and antiholomorphic Ka\v{c}-Moody currents, whereas $\mathcal{J}^3$ and $\widetilde{\mathcal{J}}^{\s 3}$ are \emph{not} holomorphic or antiholomorphic and such that $\mathcal{J}^3+\widetilde{\mathcal{J}}^{\s 3}$ and $\mathcal{J}^3-\widetilde{\mathcal{J}}^{\s 3}$ are the thermal $\mathrm{AdS_3}$ translation currents around the thermal circle and spatial circle, respectively. Note that the non-holomorphicity is because we must work in the full sine-Liouville theory; in the free linear dilaton plus $\beta\gamma$ system, $\mathcal{J}^3$ would of course be holomorphic by the equation of motion, but the sine-Liouville theory includes the $\mathcal{W}_+ + \mathcal{W}_-$ potential, described below, for which $\partial\tilde{\gamma} \neq 0$. This issue was also treated in \cite{Rangamani07} and noted earlier in \cite{Hemming01}. Below, we shall infer the correct value of $\varepsilon$ by demanding that we obtain the correct $\mathrm{AdS_3}$ Hagedorn inverse temperature \eqref{AdS Hagedorn}.

Now we may provide the explicit expressions for $\mathcal{W}_{\pm}$ for use in the $\phi\beta\gamma$ path integral. By definition, $\mathcal{W}_{\pm}$ is a weight-$(1,1)$ operator which has winding $\pm 1$ around the boundary spatial circle, no winding around the thermal circle and no momentum along either circle. In the free-field language it is given by
\begin{equation}\label{spatial winding}
\mathcal{W}_{\pm}(z,\barred{z}) = \normal{e^{\pm\frac{1}{4}\sqrt{\frac{k}{\alpha'}}[\gamma(z)+\tilde{\gamma}(\barred{z})]}e^{\mp\sqrt{\frac{k}{\alpha'}}[\int_C^z dz'\beta(z')+\int_{\barred{C}}^{\barred{z}}d\barred{z}'\tilde{\beta}(\barred{z}')]}}\normal{e^{-\sqrt{\frac{k-2}{\alpha'}}\phi(z,\barred{z})}}.
\end{equation}
Indeed, its holomorphic weight is immediately found to be
\begin{equation}
h_{\pm} = \frac{\alpha'}{2}\sqrt{\frac{k-2}{\alpha'}}\left(\frac{1}{\sqrt{\alpha'(k-2)}} - \frac{1}{2}\sqrt{\frac{k-2}{\alpha'}}\right) + \frac{\alpha'}{4}\left(\sqrt{\frac{k}{\alpha'}}\right)^2 = 1.
\end{equation}
To verify the boundary circle charges, we note that the total derivatives in \eqref{J3 nonhol} and \eqref{J3 nonantihol} are irrelevant in the untwisted sector (i.e.~for zero time winding), and we compute the OPEs
\begin{align}
\left(\beta(z) + \frac{1}{4}\partial\gamma(z)\right)\mathcal{W}_{\pm}(0) & \sim 0
\\ \frac{1}{2}\left[\gamma(z)+\tilde{\gamma}(\barred{z})\right]\mathcal{W}_{\pm}(0) & \sim \pm\sqrt{\alpha'k}\ln|z| \ \mathcal{W}_{\pm}(0)
\\ \frac{1}{2i}\left[\gamma(z)-\tilde{\gamma}(\barred{z})\right]\mathcal{W}_{\pm}(0) & \sim \mp\frac{i\sqrt{\alpha'k}}{2}\ln\left(\frac{z}{\barred{z}}\right) \mathcal{W}_{\pm}(0).
\end{align}
The first OPE implies zero momentum around both boundary circles. The second OPE implies zero winding around the thermal circle because there is no monodromy as the coordinate operator $\frac{1}{2}(\gamma+\widetilde{\gamma})$ encircles $\mathcal{W}_{\pm}$. The third OPE implies winding $\pm 1$ around the boundary spatial circle because the additive monodromy of the coordinate operator $\frac{1}{2i}(\gamma-\tilde{\gamma})$ of radius $R = \sqrt{k\alpha'}$ as $z\rightarrow e^{2\pi i}z$ is $\pm 2\pi R$. With the expression \eqref{spatial winding}, the dual $\mathrm{AdS_3}$ action in \eqref{sL action} is fully specified. As usual, the parameter $\mu$ in the action is unphysical because it can always be absorbed by a shift redefinition of the field $\phi$.

It remains to identify the vertex operators by which we would like to deform the thermal $\mathrm{AdS_3}$ CFT. To apply the Horowitz-Polchinski methodology, we are supposed to deform by the Euclidean time-winding operators which are the first states to become massless exactly at the $\mathrm{AdS_3}$ Hagedorn temperature as well as by a graviton operator which backreacts the thermal geometry in response to the presence of the winding condensate. First consider the Euclidean time-winding operators which have additive monodromy $w\beta$ as it is encircled counterclockwise by $\frac{1}{2}(\gamma+\tilde{\gamma})$ but no other boundary torus charges (as any such charges would only serve to increase the weight of the operator, and we are after the time-winding operator which first becomes massless). The general such time-winding operators and their conformal dimensions are
\begin{align}
\mathcal{W}^{w}_{j}(z,\barred{z}) & = \normal{e^{-\frac{i(1+\varepsilon)w\beta}{8\pi\alpha'}[\gamma(z)-\tilde{\gamma}(\barred{z})]}e^{\frac{iw\beta}{2\pi\alpha'}[\int_C^z dz'\beta(z')-\int_{\barred{C}}^{\barred{z}}d\barred{z}'\tilde{\beta}(\barred{z}')]}}\normal{e^{-2Qj\phi(z,\barred{z})}}
\\ \Delta^w_j & = -\frac{2j(j-1)}{k-2} - \frac{w^2(1+\varepsilon)\beta^2}{8\pi^2\alpha'},
\end{align}
because we have the OPEs
\begin{align}
\mathcal{J}^3(z,\barred{z})\mathcal{W}^w_j(0) & \sim 0
\\* \widetilde{\mathcal{J}}^{\s 3}(z,\barred{z})\mathcal{W}^w_j(0) & \sim 0
\\* \frac{1}{2}\left[\gamma(z)+\tilde{\gamma}(\barred{z})\right]\mathcal{W}^w_j(0) & \sim -\frac{iw\beta}{4\pi}\ln\left(\frac{z}{\barred{z}}\right)\mathcal{W}^w_j(0)
\\* \frac{1}{2i}\left[\gamma(z)-\tilde{\gamma}(\barred{z})\right]\mathcal{W}^w_j(0) & \sim -\frac{w\beta}{2\pi}\ln|z| \ \mathcal{W}^w_j(0).
\end{align}
The linear dilaton momentum $j$ corresponding to delta-function normalizable scattering states satisfies $j \in \frac{1}{2} + i\mathds{R}$; in the original WZW model, $j$ corresponds to the $\mathfrak{sl}(2)$ spin of the operator. In the dual description here, $\phi$ corresponds to the radial direction in the bulk, so when writing $j = \frac{1}{2} + ip$, the real quantity $p$ is essentially the $\mathrm{AdS_3}$ radial momentum of the scattering state, which will be made more precise in the following subsection. For now, we still must determine the value of the parameter $\varepsilon$ in the translation currents \eqref{J3 nonhol} and \eqref{J3 nonantihol}. The critical temperature $\beta_*$ where a state corresponding to $\mathcal{W}^w_j$ must first condense is the one for which the lightest state (i.e.~$w = \pm 1$ and $p=0$) becomes marginal at $\beta = \beta_*$. This condition reads
\begin{equation}
\Delta^{w = \pm 1}_{j=\frac{1}{2}} = \frac{1}{2(k-2)} - \frac{(1+\varepsilon)\beta_*^2}{8\pi^2\alpha'} \stackrel{!}{=} 2,
\end{equation}
which determines this critical temperature as
\begin{equation}
\beta_* = 4\pi\sqrt{-\frac{\alpha'}{1+\varepsilon}\left(1 - \frac{1}{4(k-2)}\right)}.
\end{equation}
Demanding that $\beta_*$ be the $\mathrm{AdS_3}$ Hagedorn inverse temperature $\beta_{\text{H}} = 4\pi\sqrt{\alpha'(1-\frac{1}{4(k-2)})}$ determines 
\begin{equation}
\varepsilon = -2,
\end{equation}
which we set henceforth. Note that, while $\mathcal{W}^w_j$ are neutral under the torus translation currents $\mathcal{J}^3$ and $\widetilde{\mathcal{J}}^{\s 3}$, they are charged under the holomorphic and antiholomorphic Ka\v{c}-Moody currents $J^3$ and $\widetilde{J}^{\s 3}$, namely
\begin{align}
J^3(z)\mathcal{W}^w_j(0) & \sim -\frac{iw\beta}{4\pi z}\sqrt{\frac{k}{\alpha'}} \s\s \mathcal{W}^w_j(0)
\\ \widetilde{J}^{\s 3}(\barred{z})\mathcal{W}^w_j(0) & \sim +\frac{iw\beta}{4\pi\barred{z}}\sqrt{\frac{k}{\alpha'}}\s\s \mathcal{W}^w_j(0).
\end{align}
We shall use this result later.

Finally, we need the appropriate graviton vertex operators. By the symmetries of the problem, they must be neutral under all boundary torus charges. Correspondingly, such vertex operators cannot involve exponentials of $\beta$ or $\gamma$, instead constructed solely out of the currents $J^3$ and $\widetilde{J}^{\s 3}$. The general such operators and their conformal dimensions are
\begin{align}
\mathcal{V}^0_j(z,\barred{z}) & = J^3(z)\widetilde{J}^{\s 3}(\barred{z})\normal{e^{-2Qj\phi(z,\barred{z})}}
\\* \Delta_j & = 2 - \frac{2j(j-1)}{k-2},
\end{align}
where again $j \in \frac{1}{2} + i\mathds{R}$. Note that it is the holomorphic and antiholomorphic Ka\v{c}-Moody currents which appear in the massless string vertex operators.

\subsection{Correct Basis of Operators}

While we have seen that the heuristic operators we shall need are
\begin{align}\label{heuristic operators}
\mathcal{W}^{w}_{j}(z,\barred{z}) & = \normal{e^{\frac{iw\beta}{8\pi\alpha'}[\gamma(z)-\tilde{\gamma}(\barred{z})]}e^{\frac{iw\beta}{2\pi\alpha'}[\int_C^z dz'\beta(z')-\int_{\barred{C}}^{\barred{z}}d\barred{z}'\tilde{\beta}(\barred{z}')]}}\normal{e^{-2Qj\phi(z,\barred{z})}}
\\ \Delta^w_j & = -\frac{2j(j-1)}{k-2} + \frac{w^2\beta^2}{8\pi^2\alpha'}
\end{align}
for $w = \pm 1$ and $w = 0$ (as for the latter it will be trivial to append the currents $J^3$ and $\widetilde{J}^{\s 3}$), the expressions \eqref{heuristic operators} are not the correct string vertex operators. As usual, the presence of the (sine-)Liouville wall in \eqref{sL action} couples operators with linear dilaton momenta $j$ and $1-j$, and we must identify the correct linear combination which is canonically normalized. Thus, we must compute the two-point functions $\langle\mathcal{W}^w_{j_1}(z_1,\barred{z}_1)\mathcal{W}^{-w}_{j_2}(z_2,\barred{z}_2)\rangle$ in the full sine-Liouville theory, which we do in this subsection. However, there are subtleties involved in computing two-point functions directly from the sine-Liouville path integral, so we shall instead first compute the three-point functions $\left\langle\mathcal{W}^w_{j_1}(z_1,\barred{z}_1)\mathcal{W}^{-w}_{j_2}(z_2,\barred{z}_2)\mathcal{W}^0_{j_3}(z_3,\barred{z}_3)\right\rangle$ and then take the $j_3\rightarrow 0$ limit to obtain the two-point functions. For the reader who wishes to skip the technical details of the derivation, the final result is contained in \eqref{correct basis} and \eqref{correct basis normalization}.

The standard computational trick involves performing the dilaton zero-mode integration in the path integral, choosing the parameters in the correlator such that the resulting correlator involves an integral power of the sine-Liouville operator, computing the resulting free correlators and integrals and finally obtain the desired correlators by analytic continuation back to the original values of the parameters. 

The first step is to separate out the dilaton zero-mode dependence of the operators involved by writing
\begin{align}
\mathcal{W}_{\pm} & = \mathcal{W}_{\pm}^{\prime}e^{-\frac{1}{\alpha'Q}\phi_0}
\\ \mathcal{W}^w_j & = \mathcal{W}_j^{\prime w}e^{-2Qj\phi_0},
\end{align}
where primes denote the removal of the zero-mode $\phi_0$. Next, we perform the exact zero-mode integration over the dilaton zero-mode by using the elementary integral
\begin{align}
\int_{-\infty}^{\infty}d\phi_0 \ e^{-a\phi_0 - be^{-c\phi_0}} & = \frac{1}{bc}\int_0^{\infty}d(be^{-c\phi_0}) \ \left(\frac{be^{-c\phi_0}}{b}\right)^{\frac{a}{c}-1}e^{-be^{-c\phi_0}} = \frac{1}{c}\Gamma\left(\frac{a}{c}\right)b^{-\frac{a}{c}}.
\end{align}
Writing the sine-Liouville action as
\begin{equation}
S_{\text{sL}} = S_{\text{free}} - Q\chi\phi_0 + \mu\int d^2 z\left(\mathcal{W}_+^{\prime} + \mathcal{W}_-^{\prime}\right)e^{-\frac{1}{\alpha'Q}\phi_0},
\end{equation}
we express the desired three-point function as
\begin{align}
\notag \hspace{30pt}&\hspace{-30pt}\left\langle\mathcal{W}^w_{j_1}(z_1,\barred{z}_1)\mathcal{W}^{-w}_{j_2}(z_2,\barred{z}_2)\mathcal{W}^0_{j_3}(z_3,\barred{z}_3)\right\rangle_{\text{sL}}
\\* & = \int\mathcal{D}\phi\mathcal{D}\beta\mathcal{D}\gamma \ e^{-S_{\text{sL}}}e^{-2Q(j_1+j_2+j_3)\phi_0}\mathcal{W}^{\prime w}_{j_1}(z_1,\barred{z}_1)\mathcal{W}^{\prime -w}_{j_2}(z_2,\barred{z}_2)\mathcal{W}^{\prime 0}_{j_3}(z_3,\barred{z}_3)
\\* & = \mathcal{N}_0\alpha'Q\Gamma(-2n)\hspace{-2pt}\left\langle\hspace{-2pt}\mathcal{W}^{\prime w}_{j_1}(z_1,\barred{z}_1)\mathcal{W}^{\prime -w}_{j_2}(z_2,\barred{z}_2)\mathcal{W}^{\prime 0}_{j_3}(z_3,\barred{z}_3)\hspace{-2pt}\left(\hspace{-1pt}\mu\hspace{-2pt}\int \hspace{-2pt}d^2 z(\mathcal{W}_+^{\prime} \hspace{-2pt}+\hspace{-2pt} \mathcal{W}_-^{\prime})(z,\barred{z})\hspace{-2pt}\right)^{\hspace{-2pt}2n}\right\rangle_{\hspace{-2pt}\text{free}}^{\prime}\hspace{-2pt},
\end{align}
where
\begin{equation}
n \equiv -\alpha'Q^2\left(j_1 + j_2 + j_3 - \frac{\chi}{2}\right),
\end{equation}
and $\langle\cdots\rangle_{\text{free}}^{\prime}$ indicates a correlator in the pure linear dilaton plus $\beta\gamma$ system with the dilaton zero-mode removed. The factor $\mathcal{N}_0$ appearing above is a normalization constant that appears because the zero-mode integration is performed at the level of the partition function, not the correlator which is supposed to be normalized by dividing by the bare partition function. It is the zero-mode-removed correlator $\langle\cdots\rangle^{\prime}_{\text{free}}$ which is normalized by the zero-mode-removed bare partition function, and $\mathcal{N}_0$ thus arises to ensure that the zero-mode contribution to the actual correlator is correctly normalized. Note that there is no way to compute this normalization constant $\mathcal{N}_0$ from the sine-Liouville path integral itself, but we shall eventually fix it by matching to the flat space limit.

The preceding expression makes sense as a free-field correlator when $n$ is a non-negative integer or half-integer, but the free theory has winding conservation on both circles, so $2n$ must be even for a nonvanishing result. Thus, we analytically continue the parameter $k$ so that $n \in \mathds{Z}^+$. One might be concerned that the $\Gamma(-2n)$ prefactor then diverges, but in fact we shall see that this divergence is canceled by the correlator, leaving a finite result. Using spatial winding conservation of the free correlator, we further write
\begin{multline}
\left\langle\mathcal{W}^w_{j_1}(z_1,\barred{z}_1)\mathcal{W}^{-w}_{j_2}(z_2,\barred{z}_2)\mathcal{W}^0_{j_3}(z_3,\barred{z}_3)\right\rangle_{\text{sL}} = -\frac{\pi\alpha'Q}{n!^2\sin(2\pi n)}\mu^{2n}
\\* \times \left\langle\hspace{-2pt}\mathcal{W}^{\prime w}_{j_1}(z_1,\barred{z}_1)\mathcal{W}^{\prime -w}_{j_2}(z_2,\barred{z}_2)\mathcal{W}^{\prime 0}_{j_3}(z_3,\barred{z}_3)\prod_{i=4}^{n+3}\int \hspace{-2pt}d^2 z_i\hspace{-2pt}\int \hspace{-2pt}d^2 z_{i+n} \mathcal{W}_+^{\prime}(z_i,\barred{z}_i)\mathcal{W}_-^{\prime}(z_{i+n},\barred{z}_{i+n})\hspace{-2pt}\right\rangle_{\hspace{-3pt}\text{free}}^{\hspace{-1pt}\prime}\hspace{-3pt}.
\end{multline}
The base correlator integrand is easily found using Wick contractions to be
\begin{multline}
\left\langle\mathcal{W}^{\prime w}_{j_1}(z_1,\barred{z}_1)\mathcal{W}^{\prime -w}_{j_2}(z_2,\barred{z}_2)\mathcal{W}^{\prime 0}_{j_3}(z_3,\barred{z}_3)\prod_{i=4}^{n+3}\mathcal{W}_+^{\prime}(z_i,\barred{z}_i)\mathcal{W}_-^{\prime}(z_{i+n},\barred{z}_{i+n})\right\rangle_{\text{free}}^{\prime}
\\ = \frac{1}{|z_{12}|^{\frac{w^2\beta^2}{4\pi^2\alpha'}+4\alpha' Q^2 j_1 j_2}|z_{13}|^{4\alpha' Q^2 j_1 j_3}|z_{23}|^{4\alpha' Q^2 j_2 j_3}}
\\ \times \prod_{i=4}^{2n+3}\frac{1}{|z_{1i}|^{2j_1}|z_{2i}|^{2j_2}|z_{3i}|^{2j_3}}\prod_{\substack{i,j=4 \\ i < j}}^{n+3} \left|z_{ij}z_{i+n,j+n}\right|^{2} \prod_{i,j=4}^{n+3}\frac{1}{\left|z_{i,j+n}\right|^{2(k-1)}},
\end{multline}
since $\langle\mathds{1}\rangle_{\text{free}}^{\prime} = 1$ by the definition of the zero-mode-removed correlator being normalized by the zero-mode-removed partition function. The moduli integrals can be computed in closed form using the results of \cite{Fukuda01}. Using the standard notation
\begin{equation}
\gamma(x) \equiv \frac{\Gamma(x)}{\Gamma(1-x)},
\end{equation}
the requisite integral identity is
\begin{multline}\label{integral identity}
\prod_{i=4}^{n+3}\int \hspace{-2pt}d^2 z_i \hspace{-2pt}\int \hspace{-2pt}d^2 z_{i+n}\hspace{-2pt}\prod_{i=4}^{2n+3}\frac{1}{|z_{1i}|^{2j_1}|z_{2i}|^{2j_2}|z_{3i}|^{2j_3}}\prod_{\substack{i,j=4 \\ i < j}}^{n+3} \left|z_{ij}z_{i+n,j+n}\right|^{2} \prod_{i,j=4}^{n+3}\frac{1}{\left|z_{i,j+n}\right|^{2-\frac{2(j_1+j_2+j_3-1)}{n}}}
\\ = \frac{2^{2(j_1+j_2+j_3-2)}(2\pi)^{2n}n!^2\gamma(\frac{j_1+j_2-j_3}{2})\gamma(\frac{j_1-j_2+j_3}{2})\gamma(\frac{-j_1+j_2+j_3}{2})\gamma(\frac{j_1+j_2+j_3-1}{2})}{\gamma(j_1)\gamma(j_2)\gamma(j_3)|z_{12}|^{2(j_1+j_2-j_3)n}|z_{23}|^{2(j_2+j_3-j_1)n}|z_{31}|^{2(j_3+j_1-j_2)n}}
\\ \times \prod_{\ell=1}^{n-1}\frac{\gamma\big(\frac{\ell(j_1+j_2+j_3-1)}{n}\big)}{\gamma\big(2j_1-\frac{\ell(j_1+j_2+j_3-1)}{n}\big)\gamma\big(2j_2-\frac{\ell(j_1+j_2+j_3-1)}{n}\big)\gamma\big(2j_3-\frac{\ell(j_1+j_2+j_3-1)}{n}\big)}.
\end{multline}
The final step is to analytically continue away from integer $n$, where currently the only dependence on integrality of $n$ is in the $(n-1)$-fold product of gamma functions in \eqref{integral identity}. To that end, introduce the Barnes Upsilon function $\Upsilon_b(x)$ which is defined for $0 < \mathrm{Re}\s\s x < b + \frac{1}{b}$ by
\begin{equation}\label{Upsilon definition}
\ln\Upsilon_b(x) = \int_0^{\infty}\frac{dt}{t}\left[\bigg(\frac{b+\frac{1}{b}}{2}-x\bigg)^2 e^{-t} - \frac{\sinh^2\hspace{-2pt}\big((\frac{b+\frac{1}{b}}{2}-x)\frac{t}{2}\big)}{\sinh(\frac{bt}{2})\sinh(\frac{t}{2b})}\right]
\end{equation}
and elsewhere by analytic continuation. Useful properties of the Barnes Upsilon function are summarized in Appendix \ref{Barnes}. For our purposes here, the important identities are the shift relations
\begin{align}
\Upsilon_b(x+b) & = b^{1-2bx}\gamma(bx)\Upsilon_b(x)
\\ \Upsilon_b\left(x + \frac{1}{b}\right) & = b^{\frac{2x}{b}-1}\gamma\left(\frac{x}{b}\right)\Upsilon_b(x).
\end{align}
Then, defining the parameter
\begin{equation}
b \equiv \sqrt{k-2},
\end{equation}
we may represent the relevant $(n-1)$-fold products for integer $n = -\frac{1}{b^2}(j_1+j_2+j_3-1)$ as
\begin{align}
\prod_{\ell=1}^{n-1}\gamma\left(x - \frac{\ell(j_1+j_2+j_3-1)}{n}\right) & = \frac{b^{n[2x-1+(n-1)b^2]}}{\gamma(x)}\frac{\Upsilon_b(\frac{x}{b}+nb)}{\Upsilon_b(\frac{x}{b})}
\\ \prod_{\ell=1}^{n-1}\gamma\left(\frac{\ell(j_1+j_2+j_3-1)}{n}\right) & = \frac{b^{n[2(j_1+j_2+j_3-1)-1+(n-1)b^2]}}{\gamma(j_1+j_2+j_3-1)}\frac{\Upsilon_b(\frac{j_1+j_2+j_3-1}{b}+nb)}{\Upsilon_b(\frac{j_1+j_2+j_3-1}{b})}.
\end{align}
The numerator Barnes Upsilon function in the final expression vanishes (i.e.~$\Upsilon_b(0) = 0$), which cancels against the previous $\sin(2\pi n)$ in the denominator as
\begin{equation}
\frac{\Upsilon_b(\frac{j_1+j_2+j_3-1}{b}+nb)}{\sin(2\pi n)} \stackrel{n \in \mathds{Z}^+}{\longrightarrow} \frac{b}{2\pi}\Upsilon'_b(0),
\end{equation}
where the derivative of the Barnes Upsilon function at zero is given by $\Upsilon'_b(0) = \Upsilon_b\left(\frac{1}{b}\right)$. The resulting expression does not rely on the integrality of $n$, and so the analytic continuation is complete. Therefore, we have computed this sine-Liouville sphere three-point function as
\begin{multline}
\left\langle\mathcal{W}_{j_1}^{w}(z_1,\barred{z}_1)\mathcal{W}_{j_2}^{-w}(z_2,\barred{z}_2)\mathcal{W}_{j_3}^{0}(z_3,\barred{z}_3)\right\rangle_{\text{sL}}
\\ = \frac{C^{(w,-w,0)}(j_1,j_2,j_3)}{|z_{12}|^{\Delta_{j_1}^{w}+\Delta_{j_2}^{-w}-\Delta_{j_3}^{0}}|z_{13}|^{\Delta_{j_1}^{w}-\Delta_{j_2}^{-w}+\Delta_{j_3}^{0}}|z_{23}|^{-\Delta_{j_1}^{w}+\Delta_{j_2}^{-w}+\Delta_{j_3}^{0}}},
\end{multline}
where
\begin{multline}\label{3-point not normalized}
C^{(w,-w,0)}(j_1,j_2,j_3) = -\frac{\sqrt{\alpha'}\mathcal{N}_0}{2}\hspace{-3pt}\left(\hspace{-2pt}\frac{2(4b)^{b^2\hspace{-1pt}-\hspace{-1pt}1}}{\pi\mu}\hspace{-2pt}\right)^{\hspace{-4pt}\frac{2(\hspace{-1pt}j_1\hspace{-1pt}+\hspace{-1pt}j_2\hspace{-1pt}+\hspace{-1pt}j_3\hspace{-1pt}-\hspace{-1pt}1\hspace{-1pt})}{b^2}}\hspace{-5pt}\frac{\gamma(\frac{j_1+j_2-j_3}{2})\gamma(\frac{j_1-j_2+j_3}{2})\gamma(\frac{-j_1+j_2+j_3}{2})}{\gamma(\frac{1}{2}\hspace{-2pt}-\hspace{-2pt}j_1)\gamma(\frac{1}{2}\hspace{-2pt}-\hspace{-2pt}j_2)\gamma(\frac{1}{2}\hspace{-2pt}-\hspace{-2pt}j_3)\gamma(\frac{j_1+j_2+j_3}{2})}
\\ \times \frac{\Upsilon'_b(0)\Upsilon_b(\frac{2j_1}{b})\Upsilon_b(\frac{2j_2}{b})\Upsilon_b(\frac{2j_3}{b})}{\Upsilon_b(\frac{1-(-j_1+j_2+j_3)}{b})\Upsilon_b(\frac{1-(j_1-j_2+j_3)}{b})\Upsilon_b(\frac{1-(j_1+j_2-j_3)}{b})\Upsilon_b(\frac{j_1+j_2+j_3-1}{b})}.
\end{multline}
We note that this three-point coefficient is independent of $w$ and is totally symmetric in $j_1$, $j_2$ and $j_3$. Moreover, it satisfies the reflection relation
\begin{align}
C^{(w,-w,0)}(j_1,j_2,1-j_3) & = \frac{1}{S(j_3)}C^{(w,-w,0)}(j_1,j_2,j_3)
\\ \label{reflection coefficient} S(j) & \equiv -\left(\frac{b^2}{(2\pi\mu)^{1/b^2}}\right)^{2(2j-1)}\gamma^2(1-j)\frac{\Gamma(2j-1)\Gamma(\frac{2j-1}{b^2})}{\Gamma(1-2j)\Gamma(\frac{1-2j}{b^2})}
\\ S(j)S(1-j) & = 1.
\end{align}
Furthermore, this reflection coefficient $S(j)$ agrees with that of the $\mathrm{SL}(2,\mathds{R})_k$ WZW model, as obtained for instance in \cite{Maldacena01}.

Now we may obtain the desired two-point functions by taking $j_1,j_2 \in \frac{1}{2} + i\mathds{R}$ to lie on the physical branch while taking the limit $j_3\rightarrow 0$. Due to the factor of $\Upsilon_b(\frac{2j_3}{b})$ which vanishes linearly as $j_3\rightarrow 0$, the correlator above vanishes in this limit except when the numerator gamma functions have a compensating pole or when the denominator Barnes Upsilon functions have a compensating zero. To that end, we note the limits 
\begin{equation}
\text{$j_1 \simeq j_2$:} \qquad \gamma\hspace{-2pt}\left(\frac{j_1\hspace{-2pt}-\hspace{-2pt}j_2\hspace{-2pt}+\hspace{-2pt}j_3}{2}\right)\gamma\hspace{-2pt}\left(\frac{-j_1\hspace{-2pt}+\hspace{-2pt}j_2\hspace{-2pt}+\hspace{-2pt}j_3}{2}\right) \stackrel{j_3\rightarrow 0}{\longrightarrow} \frac{4}{j_3^2-(j_1-j_2)^2}.
\end{equation}
and
\begin{equation}
\text{$j_1+j_2 \simeq 1$:} \qquad \Upsilon_b\hspace{-2pt}\left(\frac{1\hspace{-2pt}-\hspace{-2pt}(j_1\hspace{-2pt}+\hspace{-2pt}j_2\hspace{-2pt}-\hspace{-2pt}j_3)}{b}\right)\Upsilon_b\hspace{-2pt}\left(\frac{j_1\hspace{-2pt}+\hspace{-2pt}j_2\hspace{-2pt}+\hspace{-2pt}j_3\hspace{-2pt}-\hspace{-2pt}1}{b}\right) \stackrel{j_3\rightarrow 0}{\longrightarrow} \frac{j_3^2\hspace{-2pt}-\hspace{-2pt}(j_1\hspace{-2pt}+\hspace{-2pt}j_2\hspace{-2pt}-\hspace{-2pt}1)^2}{b^2}\Upsilon'_b(0)^2.
\end{equation}
Noting the appearance of the nascent delta function $\delta_{\epsilon}(x) = \frac{\epsilon}{\pi(x^2+\epsilon^2)}$, we hence conclude that the only nontrivial part of the $j_3\rightarrow 0$ limit on the physical branch $j_{1,2} \in \frac{1}{2} + i\mathds{R}$ is
\begin{multline}
\text{$j_{1,2} = \frac{1}{2} \hspace{-2pt}+\hspace{-2pt} ip_{1,2}$:} \qquad \gamma\hspace{-2pt}\left(\frac{j_1\hspace{-2pt}-\hspace{-2pt}j_2\hspace{-2pt}+\hspace{-2pt}j_3}{2}\right)\gamma\hspace{-2pt}\left(\frac{-j_1\hspace{-2pt}+\hspace{-2pt}j_2\hspace{-2pt}+\hspace{-2pt}j_3}{2}\right)\hspace{-2pt}\frac{\Upsilon_b(\frac{2j_3}{b})}{\Upsilon_b(\frac{1-(j_1+j_2-j_3)}{b})\Upsilon_b(\frac{j_1+j_2+j_3-1}{b})}
\\ \stackrel{j_3\rightarrow 0}{\longrightarrow} \frac{8\pi\Upsilon'_b(0)}{b\Upsilon_b(-\frac{2ip_1}{b})\Upsilon_b(\frac{2ip_1}{b})}\delta(p_1\hspace{-2pt}-\hspace{-2pt}p_2) + \frac{2\pi b}{\Upsilon'_b(0)p_1^2}\delta(p_1\hspace{-2pt}+\hspace{-2pt}p_2).
\end{multline}
Simplifying the remainder of \eqref{3-point not normalized} on the supports of these delta functions, we have therefore derived the sine-Liouville sphere two-point function
\begin{multline}
\left\langle\mathcal{W}_{\frac{1}{2}+ip}^{w}(z_1,\barred{z}_1)\mathcal{W}_{\frac{1}{2}+ip'}^{-w}(z_2,\barred{z}_2)\right\rangle_{\text{sL}} 
\\ = -\frac{\pi b\sqrt{\alpha'}\mathcal{N}_0}{|z_{12}|^{2\Delta^w_p}}\hspace{-2pt}\left[-\left(\hspace{-2pt}\frac{2b^2}{(2\pi\mu)^{\frac{1}{b^2}}}\hspace{-2pt}\right)^{\hspace{-2pt}4ip}\frac{\Gamma(ip)\Gamma(\frac{1}{2}\hspace{-2pt}-\hspace{-2pt}ip)\Gamma(\frac{2ip}{b^2})}{\Gamma(-ip)\Gamma(\frac{1}{2}\hspace{-2pt}+\hspace{-2pt}ip)\Gamma(-\frac{2ip}{b^2})}\delta(p\hspace{-2pt}-\hspace{-2pt}p') + \delta(p\hspace{-2pt}+\hspace{-2pt}p')\right],
\end{multline}
where $\Delta^w_p = \frac{1+4p^2}{2(k-2)} + \frac{w^2\beta^2}{8\pi^2\alpha'}$. Now define the function
\begin{equation}
\mathcal{S}(p) \equiv -\left(\frac{2b^2}{(2\pi\mu)^{\frac{1}{b^2}}}\right)^{4ip}\frac{\Gamma(ip)\Gamma(\frac{1}{2}-ip)\Gamma(\frac{2ip}{b^2})}{\Gamma(-ip)\Gamma(\frac{1}{2}+ip)\Gamma(-\frac{2ip}{b^2})},
\end{equation}
which manifestly satisfies $\mathcal{S}(p)\mathcal{S}(-p) = 1$; of course, $\mathcal{S}(p)$ is just the reflection coefficient \eqref{reflection coefficient} evaluated on the physical branch. Then, we immediately see that the combination $\frac{1}{\sqrt{\mathcal{S}(p)}}\mathcal{W}^w_{\frac{1}{2}+ip} - \sqrt{\mathcal{S}(p)}\mathcal{W}^w_{\frac{1}{2}-ip}$ is a null operator (i.e.~has vanishing two-point function with its winding-conjugate). The correct basis of operators is thus the orthogonal linear combination,
\begin{equation}\label{correct basis}
\mathcal{O}^w_p \equiv \frac{\sqrt{\alpha'}}{2\sqrt{-\mathcal{N}_0}}\left(\frac{1}{\sqrt{\mathcal{S}(p)}}\mathcal{W}^w_{\frac{1}{2}+ip} + \sqrt{\mathcal{S}(p)}\mathcal{W}^w_{\frac{1}{2}-ip}\right),
\end{equation}
with the complete set spanned by $p \in \mathds{R}^+$. The operators \eqref{correct basis} are now canonically normalized with sphere two-point function
\begin{equation}\label{correct basis normalization}
\left\langle\mathcal{O}^w_p(z_1,\barred{z}_1)\mathcal{O}^{-w}_{p'}(z_2,\barred{z}_2)\right\rangle_{\text{sL}} = \frac{\pi\alpha'\delta(p-p')}{Q|z_{12}|^{2\Delta^w_p}}.
\end{equation}

\subsection{The String Star Equations}

It is now trivial to obtain from \eqref{3-point not normalized} the three-point function of the correctly normalized string vertex operators \eqref{correct basis}. Using the duplication identity
\begin{equation}
\gamma\left(\frac{1}{2}-j\right)\gamma(1-j) = 2^{4j-1}\gamma(1-2j),
\end{equation}
we find that the winding-winding-tachyon three-point function is
\begin{equation}
\left\langle \hspace{-2pt}\mathcal{O}^w_{j_1}\hspace{-1pt}(\hspace{-1pt}z_1,\hspace{-1pt}\barred{z}_1\hspace{-1pt})\mathcal{O}^{-w}_{j_2}\hspace{-1pt}(\hspace{-1pt}z_2,\hspace{-1pt}\barred{z}_2\hspace{-1pt})\mathcal{O}^{0}_{j_3}\hspace{-1pt}(\hspace{-1pt}z_3,\hspace{-1pt}\barred{z}_3\hspace{-1pt})\hspace{-2pt}\right\rangle = \frac{\mathcal{C}(j_1,j_2,j_3)}{|z_{12}|^{\Delta_{j_1}^{w}\hspace{-1pt}+\Delta^{-w}_{j_2}\hspace{-1pt}-\Delta^{0}_{j_3}}|z_{13}|^{\Delta^{w}_{j_1}\hspace{-1pt}-\Delta^{-w}_{j_2}\hspace{-1pt}+\Delta^{0}_{j_3}}|z_{23}|^{-\Delta^{w}_{j_1}\hspace{-1pt}+\Delta^{-w}_{j_2}\hspace{-1pt}+\Delta^{0}_{j_3}}},
\end{equation}
where
\begin{multline}\label{3-point normalized}
\mathcal{C}(j_1,j_2,j_3) = \frac{b^2\alpha'^2}{4\sqrt{-\mathcal{N}_0}}\left(\frac{1}{2\pi\mu}\right)^{\frac{1}{b^2}}b^{2(1+\frac{1}{b^2})(1\hspace{-1pt}-\hspace{-1pt}j_1\hspace{-1pt}-\hspace{-1pt}j_2\hspace{-1pt}-\hspace{-1pt}j_3)}\prod_{\ell=1}^3\sqrt{-\frac{\Gamma(1-2j_{\ell})\Gamma(\frac{1-2j_{\ell}}{b^2})}{\Gamma(2j_{\ell}-1)\Gamma(\frac{2j_{\ell}-1}{b^2})}}
\\ \times \frac{\gamma\hspace{-1pt}(\hspace{-2pt}\frac{1\hspace{-1pt}+\hspace{-1pt}j_1\hspace{-1pt}-\hspace{-1pt}j_2\hspace{-1pt}-\hspace{-1pt}j_3}{2}\hspace{-2pt})\hspace{-1pt}\gamma\hspace{-1pt}(\hspace{-2pt}\frac{1\hspace{-1pt}-\hspace{-1pt}j_1\hspace{-1pt}+\hspace{-1pt}j_2\hspace{-1pt}-\hspace{-1pt}j_3}{2}\hspace{-2pt})\hspace{-1pt}\gamma\hspace{-1pt}(\hspace{-2pt}\frac{1\hspace{-1pt}-\hspace{-1pt}j_1\hspace{-1pt}-\hspace{-1pt}j_2\hspace{-1pt}+\hspace{-1pt}j_3}{2}\hspace{-2pt})\hspace{-1pt}\gamma\hspace{-1pt}(\hspace{-2pt}\frac{j_1\hspace{-1pt}+\hspace{-1pt}j_2\hspace{-1pt}+\hspace{-1pt}j_3\hspace{-1pt}-\hspace{-1pt}1}{2}\hspace{-2pt})\Upsilon'_{b}(0)\Upsilon_b(\frac{2j_1}{b})\Upsilon_b(\frac{2j_2}{b})\Upsilon_b(\frac{2j_3}{b})}{\gamma(1\hspace{-2pt}-\hspace{-2pt}2j_1)\gamma(1\hspace{-2pt}-\hspace{-2pt}2j_2)\gamma(1\hspace{-2pt}-\hspace{-2pt}2j_3)\Upsilon_{\hspace{-1pt}b}(\hspace{-1pt}\frac{\hspace{-1pt}-\hspace{-1pt}j_1\hspace{-1pt}+\hspace{-1pt}j_2\hspace{-1pt}+\hspace{-1pt}j_3}{b}\hspace{-1pt})\hspace{-1pt}\Upsilon_{\hspace{-1pt}b}\hspace{-1pt}(\hspace{-1pt}\frac{j_1\hspace{-1pt}-\hspace{-1pt}j_2\hspace{-1pt}+\hspace{-1pt}j_3}{b}\hspace{-1pt})\hspace{-1pt}\Upsilon_{\hspace{-1pt}b}\hspace{-1pt}(\hspace{-1pt}\frac{j_1\hspace{-1pt}+\hspace{-1pt}j_2\hspace{-1pt}-\hspace{-1pt}j_3}{b}\hspace{-1pt})\hspace{-1pt}\Upsilon_{\hspace{-1pt}b}\hspace{-1pt}(\hspace{-1pt}\frac{2\hspace{-1pt}-\hspace{-1pt}(\hspace{-1pt}j_1\hspace{-1pt}+\hspace{-1pt}j_2\hspace{-1pt}+\hspace{-1pt}j_3\hspace{-1pt})}{b}\hspace{-1pt})}.
\end{multline}
Now we would like to deform the thermal $\mathrm{AdS_3}$ background near the Hagedorn temperature by the winding and transverse graviton vertex operators\footnote{The minus sign in $\mathcal{O}^{(0)}_p$ is due to the fact that it is the operator $i\sqrt{\frac{2}{k}}J^3$ which has unit two-point coefficient.}
\begin{align}
\mathcal{O}^{\pm}_p(z,\barred{z}) & \equiv \mathcal{O}^{w=\pm 1}_{j=\frac{1}{2}+ip}(z,\barred{z})
\\ \mathcal{O}^{(0)}_p(z,\barred{z}) & \equiv -\frac{2}{k}J^3(z)\widetilde{J}^{\s 3}(\barred{z})\mathcal{O}^{w=0}_{j=\frac{1}{2}+ip}(z,\barred{z}).
\end{align}
The notation $\mathcal{O}_p^{(0)}$ is to remind us that this vertex operator describes a massless string scattering state. Specifically, we consider the deformed theory
\begin{equation}
S = S_{\mathrm{AdS_3}(\beta)} + \int d^2 z \int_0^{\infty}\frac{dp}{2\pi}\left[f^-(p)\mathcal{O}^+_p(z,\barred{z}) + f^+(p)\mathcal{O}^-_p(z,\barred{z}) + f^{(0)}(p)\mathcal{O}^{(0)}_p(z,\barred{z})\right] + \dotsc,
\end{equation}
where the ``$+\dotsc$'' denotes the inclusion of the other string vertex operators which we would like to be subleading. As before, we let
\begin{equation}
\epsilon \equiv \frac{\beta-\beta_{\text{H}}}{\beta_{\text{H}}}
\end{equation}
parametrize the deviation from the $\mathrm{AdS_3}$ Hagedorn temperature \eqref{AdS Hagedorn}. The only nonvanishing three-point function of $\mathcal{O}^+_p$ and $\mathcal{O}^-_p$ with a massless string state is with the transverse graviton mode $\mathcal{O}^{(0)}_p$, and the three-graviton vertex here vanishes because $J^3$ is a $\mathrm{U}(1)$ current. Thus, the beta functionals for the coupling profiles $f^+(p) = f^-(p)^*$ and $f^{(0)}(p)$ to this order are
\begin{align}\label{beta winding}
\beta^{\pm}(p) & = (2-\Delta^{\pm}_{p})f^{\pm}(p) + 8\pi\alpha' Q^2\int_0^{\infty}\frac{dp'}{2\pi}\int_0^{\infty}\frac{dp''}{2\pi}\hat{\mathcal{C}}^{(0)}(p,p',p'')f^{\pm}(p')f^{(0)}(p'') + \dotsc
\\ \label{beta graviton} \beta^{(0)}(p) & = (2-\Delta_{p}^{(0)})f^{(0)}(p) + 8\pi\alpha' Q^2\int_0^{\infty}\frac{dp'}{2\pi}\int_0^{\infty}\frac{dp''}{2\pi}\hat{\mathcal{C}}^{(0)}(p,p',p'')f^+(p')f^-(p'') + \dotsc,
\end{align}
where
\begin{align}
\Delta^{\pm}_p & = 2 + \frac{2p^2}{k-2} + (4\epsilon+2\epsilon^2)\left(1-\frac{1}{4(k-2)}\right)
\\ \Delta^{(0)}_p & = 2 + \frac{2p^2}{k-2} + \frac{1}{2(k-2)}
\end{align}
and
\begin{equation}\label{ww0}
\hat{\mathcal{C}}^{(0)}(p_1,p_2,p_3) \equiv -\frac{\beta^2}{8\pi^2\alpha'}\frac{\mathcal{C}\left(\frac{1}{2}+ip_1,\frac{1}{2}+ip_2,\frac{1}{2}+ip_3\right)}{\alpha'^{3/2}};
\end{equation}
the prefactor here arises from the charges of $\mathcal{O}^{\pm}_p$ under the Ka\v{c}-Moody current $J^3$, and the extra $\sqrt{\alpha'}$ was explicitly factored out so that $\hat{\mathcal{C}}$ is dimensionless. For a fixed $k$ and $\epsilon$, the $\mathrm{AdS_3}$ string star, if it exists, is given by a nontrivial solution to the equations obtained by setting all the beta functionals to zero. It is clear that the near-Hagedorn limit $\epsilon \ll 1$ is required in order for the higher-winding contributions to be parametrically suppressed. When the massive string states are also suppressed, the winding condensate characterizing the string star is obtained by solving the cubic integral equation
\begin{multline}\label{string star equation 1}
0 \stackrel{!}{=} -\left[\frac{p^2}{k-2} + (2\epsilon + \epsilon^2)\left(1-\frac{1}{4(k-2)}\right)\right]f^{\pm}(p)
\\ + \frac{(8\pi)^2}{k\hspace{-2pt}-\hspace{-2pt}2}\hspace{-3pt}\int_0^{\infty}\hspace{-3pt}\frac{dp'}{2\pi}\hspace{-3pt}\int_0^{\infty}\hspace{-3pt}\frac{dp_1}{2\pi}\hspace{-3pt}\int_0^{\infty}\hspace{-3pt}\frac{dp_2}{2\pi}\hspace{-3pt}\int_0^{\infty}\hspace{-3pt}\frac{dp_3}{2\pi}\frac{\hat{\mathcal{C}}^{(0)}\hspace{-1pt}(p,p_1,p')\hat{\mathcal{C}}^{(0)}\hspace{-1pt}(p',p_2,p_3)}{4p'^2+1}f^{\pm}(p_1)f^+(p_2)f^-(p_3).
\end{multline}
After one obtains a solution to the preceding nonlinear integral equation, the transverse graviton backreaction is then determined algebraically as
\begin{equation}
f^{(0)}(p) = \frac{16\pi}{4p^2+1}\int_0^{\infty}\frac{dp'}{2\pi}\int_0^{\infty}\frac{dp''}{2\pi}\hat{\mathcal{C}}^{(0)}(p,p',p'')f^+(p')f^-(p'').
\end{equation}
We shall provide evidence below for the existence of these $\mathrm{AdS_3}$ string stars essentially by numerically solving \eqref{string star equation 1} in a $\frac{1}{k}$ expansion. While the equation \eqref{string star equation 1} itself is strictly valid only when the massive string states decouple as $k\rightarrow \infty$, the idea is to expand the solution in powers of $\frac{1}{k-2}$ and, as $k$ is lowered from $k \gg 1$ (the semiclassical r\'{e}gime) to $k-3 \simeq \mathcal{O}(1)$ (the very stringy r\'{e}gime), to take into account the massive string tower one by one when their presence nontrivially corrects the winding condensate profile, as deduced numerically. In this way, we propose a controlled procedure for understanding the $\mathrm{AdS_3}$ near-Hagedorn string stars well away from the semiclassical limit.

Before proceeding, let us make one final comment about the structure of the string star equations presented here. For notational clarity, revert momentarily to the discrete operator notation of \eqref{beta function compact}. Letting $\lambda^i_w$ denote the couplings of the winding deformation and $\lambda^i$ denote the couplings for the other parts of the string spectrum, we have seen that the leading-order winding beta functions take the form
\begin{align}
\beta_w^i & = (2-\Delta_i^w)\lambda^i_w + 2\pi\sum_{j,k}C_{jk}^{\phantom{jk}i}\lambda_w^j\lambda^k + \dotsc 
\\ \label{winding beta function general} & = (2-\Delta_i^w)\lambda^i_w + (2\pi)^2\sum_{j,\ell,m}\left(\sum_k\frac{C_{jk}^{\phantom{jk}i}C_{\ell m}^{\phantom{\ell m}k}}{\Delta_k-2}\right)\lambda_w^j\lambda_w^{\ell}\lambda_w^m + \dotsc,
\end{align}
where the second line results from substituting the solution to the equation $\beta^k = 0$ for the non-winding string spectrum. Thanks to being near the Hagedorn temperature, $2 - \Delta_i^w$ is $\mathcal{O}(\frac{1}{k-2})$, and $\Delta_k-2$ is also $\mathcal{O}(\frac{1}{k-2})$ only for the massless string spectrum. As such, the winding and graviton profiles are of the same order, and the profiles for the massive string deformations at level $n$ are suppressed by $\frac{1}{n(k-2)}$. There are of course other cubic terms in \eqref{winding beta function general} contained in the ``$+\dotsc$'', but they contain one winding coupling and two non-winding couplings, and are hence suppressed by $\frac{1}{k-2}$, by $\frac{1}{(k-2)^2}$ or by $\frac{1}{(k-3)^3}$ depending on whether two, one or none of the non-winding couplings are the massless graviton, respectively. Then, there are other terms in the $\frac{1}{k-2}$ expansion simply given by the cubic terms in all of the beta functions, then the quartic terms, and so on. In this way, one has a controlled expansion away from the semiclassical limit. From a string field theory perspective, the cubic term written in \eqref{winding beta function general} represents the contribution to the four-winding vertex from two three-point vertices connected by an off-shell propagator (with the denominator essentially being the $\frac{1}{L_0}$ string field propagator). The other terms at cubic order that arise from the innate cubic terms in the beta function represent the pure winding four-point vertex itself. In principle, the cubic coefficient in \eqref{winding beta function general} is fixed entirely by symmetry, but it is still a nontrivial task to compute the sum; see the first several contributions in Section \ref{stringy} below. In fact, it seems that the simplest way to incorporate the leading correction from the entire massive string tower is simply to start with the winding beta function expanded to third order, obtaining its coefficient from the full winding four-point function which already subsumes the second term in \eqref{winding beta function general}.

\subsection{Flat Space Limit}

First, we perform a nontrivial check on the preceding string star equations by taking their flat space limit. While there do not exist normalizable string star solutions above flat $S^1_{\beta}\times\mathds{R}^2$, the form of the equations must still match the $d=2$ special case of the flat space analysis performed in Section \ref{flat space}.

The flat space limit corresponds to
\begin{equation}
k \rightarrow \infty, \qquad j = \frac{1}{2} + \frac{iq}{2}\sqrt{k-2}, \quad \text{with} \ q \in \mathds{R}^+ \ \text{held fixed}.
\end{equation}
Heuristically, this is because the linear dilaton exponentials in the vertex operators are of the form $\normal{e^{-2Qj\phi}}$ so that when writing $j = \frac{1}{2} + ip$, the quantity $p$ is the dimensionless AdS radial momentum (since $Q = \frac{1}{\sqrt{\alpha'(k-2)}} = \frac{1}{\ell_{\mathrm{AdS}}}$); when $p$ itself scales like $\sqrt{k-2}$, the length scale probed by the vertex operator is $\frac{1}{Q\sqrt{k-2}} = \ell_{\text{s}}$. Thus, in writing $j = \frac{1}{2} + \frac{iq}{2}\sqrt{k-2}$, the quantity $q$ is the flat space dimensionless radial momentum used in Section \ref{flat space}. To compute the flat space limit of the three-point coefficient \eqref{3-point normalized}, we first need the $b\rightarrow \infty$ behavior of the Barnes Upsilon functions involved; the full asymptotic expansions of the Barnes Upsilon functions are contained in Appendix \ref{Barnes}. Here, we need $\Upsilon_b(\frac{2j}{b})$ and $\Upsilon_b(\frac{j}{b})$ for $j = \frac{1}{2} + \frac{iqb}{2}$ with $q$ fixed and finite as well as $\Upsilon_b'(0)$. These expansions are given by\footnote{In many places in the literature, the parameter $b$ is taken to be in the range $0 < b < 1$ instead of $1 < b < \infty$ as we take here. The expressions are equivalent because the Barnes Upsilon function obeys $\Upsilon_b(x) = \Upsilon_{1/b}(x)$.}
\begin{align}
\Upsilon_b\hspace{-2pt}\left(\hspace{-1pt}\frac{2j}{b}\hspace{-1pt}\right) & \stackrel{b\rightarrow \infty}{\sim} \frac{\sqrt{2\pi}}{\Gamma(1\hspace{-1pt}+\hspace{-1pt}iqb)}b^{\frac{b^2}{4}+iqb+\frac{1}{2}-q^2 + \frac{iq}{b} + \frac{1}{4b^2}}e^{-b^2(3\ln A-\frac{1}{12}\ln 2)}e^{\gamma_{\text{E}}q^2}\bigg[1 - \frac{i\gamma_{\text{E}}q}{b} + \mathcal{O}\left(\frac{1}{b^2}\right)\bigg]
\\ \Upsilon_b\hspace{-2pt}\left(\frac{j}{b}\right) & \stackrel{b\rightarrow \infty}{\sim} \frac{\sqrt{2\pi}}{\Gamma(\frac{1}{2}+\frac{iqb}{2})}b^{\frac{b^2}{4}+\frac{iqb}{2}-\frac{q^2}{4}}e^{-b^2(3\ln A-\frac{1}{12}\ln 2)}e^{\frac{\gamma_{\text{E}}q^2}{4}}\bigg[1 + \mathcal{O}\left(\frac{1}{b^2}\right)\bigg]
\\ \Upsilon'_b(0) & \stackrel{b\rightarrow\infty}{\sim} \sqrt{2\pi} \ b^{\frac{b^2}{4} + \frac{1}{2} + \frac{1}{4b^2}}e^{-b^2(3\ln A-\frac{1}{12}\ln 2)}\left[1 + \mathcal{O}\left(\frac{1}{b^2}\right)\right],
\end{align}
where $\gamma_{\text{E}} = 0.577215\dotsc$ is the Euler-Mascheroni constant, $A = 1.282427\dotsc$ is Glaisher's constant and the notation `$\sim$' means `is asymptotic to'. Using these expressions as well as the identity
\begin{equation}
\frac{\Gamma(x)}{\gamma(\frac{1+x}{2})} = \frac{2^{x-1}\Gamma(\frac{x}{2})\Gamma(\frac{1-x}{2})}{\sqrt{\pi}},
\end{equation}
we compute
\begin{multline}
\frac{\gamma\hspace{-1pt}(\hspace{-2pt}\frac{1\hspace{-1pt}+\hspace{-1pt}j_1\hspace{-1pt}-\hspace{-1pt}j_2\hspace{-1pt}-\hspace{-1pt}j_3}{2}\hspace{-2pt})\hspace{-1pt}\gamma\hspace{-1pt}(\hspace{-2pt}\frac{1\hspace{-1pt}-\hspace{-1pt}j_1\hspace{-1pt}+\hspace{-1pt}j_2\hspace{-1pt}-\hspace{-1pt}j_3}{2}\hspace{-2pt})\hspace{-1pt}\gamma\hspace{-1pt}(\hspace{-2pt}\frac{1\hspace{-1pt}-\hspace{-1pt}j_1\hspace{-1pt}-\hspace{-1pt}j_2\hspace{-1pt}+\hspace{-1pt}j_3}{2}\hspace{-2pt})\hspace{-1pt}\gamma\hspace{-1pt}(\hspace{-2pt}\frac{j_1\hspace{-1pt}+\hspace{-1pt}j_2\hspace{-1pt}+\hspace{-1pt}j_3\hspace{-1pt}-\hspace{-1pt}1}{2}\hspace{-2pt})\Upsilon_b'(0)\Upsilon_b(\frac{2j_1}{b})\Upsilon_b(\frac{2j_2}{b})\Upsilon_b(\frac{2j_3}{b})}{\gamma(1\hspace{-2pt}-\hspace{-2pt}2j_1)\gamma(1\hspace{-2pt}-\hspace{-2pt}2j_2)\gamma(1\hspace{-2pt}-\hspace{-2pt}2j_3)\Upsilon_{\hspace{-1pt}b}(\hspace{-1pt}\frac{\hspace{-1pt}-\hspace{-1pt}j_1\hspace{-1pt}+\hspace{-1pt}j_2\hspace{-1pt}+\hspace{-1pt}j_3}{b}\hspace{-1pt})\hspace{-1pt}\Upsilon_{\hspace{-1pt}b}\hspace{-1pt}(\hspace{-1pt}\frac{j_1\hspace{-1pt}-\hspace{-1pt}j_2\hspace{-1pt}+\hspace{-1pt}j_3}{b}\hspace{-1pt})\hspace{-1pt}\Upsilon_{\hspace{-1pt}b}\hspace{-1pt}(\hspace{-1pt}\frac{j_1\hspace{-1pt}+\hspace{-1pt}j_2\hspace{-1pt}-\hspace{-1pt}j_3}{b}\hspace{-1pt})\hspace{-1pt}\Upsilon_{\hspace{-1pt}b}\hspace{-1pt}(\hspace{-1pt}\frac{2\hspace{-1pt}-\hspace{-1pt}(\hspace{-1pt}j_1\hspace{-1pt}+\hspace{-1pt}j_2\hspace{-1pt}+\hspace{-1pt}j_3\hspace{-1pt})}{b}\hspace{-1pt})}
\\ \stackrel{b\rightarrow\infty}{\sim} \frac{\Gamma(\frac{1}{4}+\frac{i(\pm q_1 \pm q_2 \pm q_3)b}{4})b^{i(q_1+q_2+q_3)b + 2 + \frac{i(q_1+q_2+q_3)}{b} + \frac{1}{b^2}}}{(2\pi)^2 \Gamma(-iq_1 b)\Gamma(-iq_2 b)\Gamma(-iq_3 b)}\bigg[1 - \frac{i\gamma_{\text{E}}(q_1\hspace{-2pt}+\hspace{-2pt}q_2\hspace{-2pt}+\hspace{-2pt}q_3)}{b} + \mathcal{O}\left(\frac{1}{b^2}\right)\bigg],
\end{multline}
where the notation $\Gamma(\frac{1}{4}+\frac{i(\pm q_1 \pm q_2 \pm q_3)b}{4})$ means the product of all eight terms corresponding to the different choices of the signs. Thus, the asymptotic expansion for the exact winding-winding-tachyon three-point coefficient \eqref{3-point normalized} so far reads
\begin{multline}
\mathcal{C}(j_1,j_2,j_3) \stackrel{b\rightarrow\infty}{\sim} \frac{b^3\alpha'^2}{(4\pi)^2\sqrt{-\mathcal{N}_0}}\left(\frac{1}{2\pi\mu}\right)^{\frac{1}{b^2}}\left(\prod_{\ell=1}^3\sqrt{-\frac{\Gamma(\frac{-iq_{\ell}}{b})}{\Gamma(\frac{iq_{\ell}}{b})}}\right)\frac{\Gamma(\frac{1}{4}+\frac{i(\pm q_1 \pm q_2 \pm q_3)b}{4})}{|\Gamma(iq_1 b)\Gamma(iq_2 b)\Gamma(iq_3 b)|}
\\ \times \bigg[1 - \frac{i\gamma_{\text{E}}(q_1\hspace{-2pt}+\hspace{-2pt}q_2\hspace{-2pt}+\hspace{-2pt}q_3)}{b} + \mathcal{O}\left(\frac{1}{b^2}\right)\bigg].
\end{multline}
It remains to expand the Gamma functions. The expansion of the three-fold product in parentheses above is $1 + \frac{i\gamma_{\text{E}}(q_1+q_2+q_3)}{b} + \mathcal{O}(\frac{1}{b^2})$, which merely cancels the subleading $\frac{1}{b}$ term in brackets. The modulus in the denominator is given exactly by
\begin{equation}
\frac{1}{\left|\Gamma(iq_1 b)\Gamma(iq_2 b)\Gamma(iq_3 b)\right|} = \sqrt{\frac{q_1 q_2 q_3 b^3\sinh(\pi q_1 b)\sinh(\pi q_2 b)\sinh(\pi q_3 b)}{\pi^3}}.
\end{equation}
For the product of Gamma functions in the numerator, we use the Stirling series in the form
\begin{equation}
\Gamma(\tfrac{1}{4}+ix)\Gamma(\tfrac{1}{4}-ix) \stackrel{|x|\rightarrow\infty}{\sim} \frac{2\pi}{\sqrt{|x|}}e^{-\pi|x|}\left[1 + \frac{1}{64x^2} + \mathcal{O}\left(\frac{1}{x^4}\right)\right].
\end{equation}
As such, we have the expansion
\begin{multline}
\frac{\Gamma(\frac{1}{4}+\frac{i(\pm q_1 \pm q_2 \pm q_3)b}{4})}{\left|\Gamma(iq_1 b)\Gamma(iq_2 b)\Gamma(iq_3 b)\right|}
\\ \stackrel{b\rightarrow\infty}{\sim} \frac{(8\pi)^2\sqrt{2\pi q_1 q_2 q_3}\s\s e^{\frac{\pi}{4}(q_1+q_2+q_3)b-\frac{\pi}{4}|-q_1+q_2+q_3|b-\frac{\pi}{4}|q_1-q_2+q_3|b-\frac{\pi}{4}|q_1+q_2-q_3|b}}{\sqrt{b(q_1\hspace{-1pt}+\hspace{-1pt}q_2\hspace{-1pt}+\hspace{-1pt}q_3)|\hspace{-2pt}-\hspace{-1pt}q_1\hspace{-1pt}+\hspace{-1pt}q_2\hspace{-1pt}+\hspace{-1pt}q_3||q_1\hspace{-1pt}-\hspace{-1pt}q_2\hspace{-1pt}+\hspace{-1pt}q_3||q_1\hspace{-1pt}+\hspace{-1pt}q_2\hspace{-1pt}-\hspace{-1pt}q_3|}}\left[1 + \mathcal{O}\left(\frac{1}{b^2}\right)\right].
\end{multline}
Finally, we note the limit
\begin{equation}
e^{\frac{\pi}{4}(q_1+q_2+q_3)b-\frac{\pi}{4}|-q_1+q_2+q_3|b-\frac{\pi}{4}|q_1-q_2+q_3|b-\frac{\pi}{4}|q_1+q_2-q_3|b} \ \ \stackrel{b\rightarrow\infty}{\sim} \ \ \theta\big(\Delta(q_1,q_2,q_3)\big),
\end{equation}
enforcing the triangle inequality up to exponentially small corrections. Therefore, the flat space limit of the winding-winding-tachyon three-point coefficient \eqref{3-point normalized} is
\begin{equation}
\mathcal{C}(j_1,j_2,j_3) \stackrel{b\rightarrow\infty}{\sim} \frac{4b^{5/2}\alpha'^2}{\sqrt{-\mathcal{N}_0}}\hspace{-2pt}\left(\hspace{-1pt}\frac{1}{2\pi\mu}\hspace{-1pt}\right)^{\frac{1}{b^2}}\hspace{-4pt}\frac{\sqrt{2\pi q_1 q_2 q_3} \ \theta\big(\Delta(q_1,q_2,q_3)\big)\left[1 + \mathcal{O}\left(\frac{1}{b^2}\right)\right]}{\sqrt{(q_1\hspace{-2pt}+\hspace{-2pt}q_2\hspace{-2pt}+\hspace{-2pt}q_3)(\hspace{-1pt}-q_1\hspace{-2pt}+\hspace{-2pt}q_2\hspace{-2pt}+\hspace{-2pt}q_3)(q_1\hspace{-2pt}-\hspace{-2pt}q_2\hspace{-2pt}+\hspace{-2pt}q_3)(q_1\hspace{-2pt}+\hspace{-2pt}q_2\hspace{-2pt}-\hspace{-2pt}q_3)}}.
\end{equation}
In Section \ref{flat space}, we derived the asymptotically $S^1_{\beta}\times \mathds{R}^2$ string star equations
\begin{align}
0 & \stackrel{!}{=} -\left(q^2 + \frac{4(\beta^2-\beta_{\text{H}}^2)}{\beta_{\text{H}}^2}\right)g^{\pm}(q) + 8\pi\int_0^{\infty}\frac{dq'}{2\pi}\int_0^{\infty}\frac{dq''}{2\pi}\mathcal{C}_{q,q',q''}g^{\pm}(q')g(q'')
\\ 0 & \stackrel{!}{=} -q^2 g(q) + 8\pi\int_0^{\infty}\frac{dq'}{2\pi}\int_0^{\infty}\frac{dq''}{2\pi}\mathcal{C}_{q,q',q''}g^+(q')g^-(q''),
\end{align}
where
\begin{equation}
\mathcal{C}_{q_1,q_2,q_3} = -\frac{\beta^{3/2}\sqrt{q_1 q_2 q_3}\s\s\theta\big(\Delta(q_1,q_2,q_3)\big)}{2\pi^2\alpha'^{3/4}\sqrt{(q_1\hspace{-2pt}+\hspace{-2pt}q_2\hspace{-2pt}+\hspace{-2pt}q_3)(\hspace{-1pt}-q_1\hspace{-2pt}+\hspace{-2pt}q_2\hspace{-2pt}+\hspace{-2pt}q_3)(q_1\hspace{-2pt}-\hspace{-2pt}q_2\hspace{-2pt}+\hspace{-2pt}q_3)(q_1\hspace{-2pt}+\hspace{-2pt}q_2\hspace{-2pt}-\hspace{-2pt}q_3)}};
\end{equation}
by convention the deformations were of the form $\int_0^{\infty}\frac{dq}{2\pi}g(q)\mathcal{O}_q$, where the operators were normalized by $\langle\mathcal{O}^w_q(0)\mathcal{O}^{-w}_{q'}(1)\rangle = 2\pi\alpha'^{3/2}\delta(q-q')$. Here, the deformations are of the form $\frac{b}{2}\int_0^{\infty}\frac{dq}{2\pi}f(\tfrac{qb}{2})\mathcal{O}_q$ with the operators again normalized by $\langle\mathcal{O}^w_p(0)\mathcal{O}^{-w}_{p'}(1)\rangle = 2\pi\alpha'^{3/2}\delta(q-q')$, and hence the relation between the profile functions is $f_{\mathrm{AdS}}(\tfrac{qb}{2}) = \frac{2}{b}g_{\text{flat}}(q)$. With this replacement, we have thus found the flat space limit of the $\mathrm{AdS_3}$ string star equations to be
\begin{align}
0 & \stackrel{!}{=} -\left(q^2 + \frac{4(\beta^2-\beta_{\text{H}}^2)}{\beta_{\text{H}}^2}\right)g^{\pm}(q) + 8\pi\int_0^{\infty}\frac{dq'}{2\pi}\int_0^{\infty}\frac{dq''}{2\pi}\hat{\mathcal{C}}^{(0)}_{q,q',q''}g^{\pm}(q')g^{(0)}(q'')
\\ 0 & \stackrel{!}{=} -q^2 g^{(0)}(q) + 8\pi\int_0^{\infty}\frac{dq'}{2\pi}\int_0^{\infty}\frac{dq''}{2\pi}\hat{\mathcal{C}}^{(0)}_{q,q',q''}g^{+}(q')g^{-}(q''),
\end{align}
where
\begin{equation}
\hat{\mathcal{C}}^{(0)}_{q_1,q_2,q_3} = -\frac{\beta^2 b^{3/2}\sqrt{\alpha'}}{\sqrt{2\pi^3}\sqrt{-\mathcal{N}_0}}\hspace{-2pt}\left(\hspace{-1pt}\frac{1}{2\pi\mu}\hspace{-1pt}\right)^{\frac{1}{b^2}}\hspace{-4pt}\frac{\sqrt{q_1 q_2 q_3} \s\s \theta\big(\Delta(q_1,q_2,q_3)\big)}{\sqrt{(q_1\hspace{-2pt}+\hspace{-2pt}q_2\hspace{-2pt}+\hspace{-2pt}q_3)(\hspace{-1pt}-q_1\hspace{-2pt}+\hspace{-2pt}q_2\hspace{-2pt}+\hspace{-2pt}q_3)(q_1\hspace{-2pt}-\hspace{-2pt}q_2\hspace{-2pt}+\hspace{-2pt}q_3)(q_1\hspace{-2pt}+\hspace{-2pt}q_2\hspace{-2pt}-\hspace{-2pt}q_3)}}.
\end{equation}
Therefore, we find exact agreement between these sets of string star equations which, even though they do not possess nontrivial normalizable solutions, unambiguously determine the sine-Liouville path integral normalization 
\begin{equation}\label{normalization constant}
\sqrt{-\mathcal{N}_0} = \sqrt{2\pi\beta b^3\alpha'}\left(\frac{1}{2\pi\mu}\right)^{\frac{1}{b^2}}.
\end{equation}

\subsection{Semiclassical Limit}

The semiclassical limit is of the most interest, which corresponds to $k\rightarrow\infty$ and $j = \frac{1}{2} + ip$ with $p \in \mathds{R}^+$ held fixed. With this scaling, the size of AdS is parametrically larger than the string scale, and states labeled by $p$ of order unity probe the AdS length; hence the finite curvature remains important, which is what allows the string star to exist. Fortunately, most of the relevant computations were performed in the preceding subsection, as only the subleading terms in the Barnes Upsilon expansions differ. The asymptotics we need now are
\begin{align}
\Upsilon_b\left(\frac{x}{b}\right) & \stackrel{b\rightarrow \infty}{\sim} \frac{\sqrt{2\pi}}{\Gamma(x)}b^{\frac{b^2}{4} + \frac{2x-1}{2} 
+ \frac{(2x-1)^2}{4b^2}}e^{-b^2(3\ln A - \frac{1}{12}\ln 2)}\left[1 - \left(\frac{(2x\hspace{-2pt}-\hspace{-2pt}1)^2}{4}\gamma_{\text{E}} - a\right)\hspace{-2pt}\frac{1}{b^2} + \mathcal{O}\hspace{-2pt}\left(\frac{1}{b^4}\right)\right]
\\ \Upsilon'_b(0) & \stackrel{b\rightarrow\infty}{\sim} \sqrt{2\pi} \ b^{\frac{b^2}{4} + \frac{1}{2} + \frac{1}{4b^2}}e^{-b^2(3\ln A - \frac{1}{12}\ln 2)}\left[1 - \left(\frac{\gamma_{\text{E}}}{4} - a\right)\frac{1}{b^2} + \mathcal{O}\left(\frac{1}{b^4}\right)\right].
\end{align}
where $a \equiv \frac{1}{12}\big(\gamma_{\text{E}}+\frac{\Gamma'(\frac{1}{2})}{\sqrt{\pi}}\big)$. Then, the asymptotic expansion for the exact winding-winding-tachyon three-point coefficient \eqref{3-point normalized} is
\begin{equation}\label{3-point normalized semiclassical}
\mathcal{C}(j_1,j_2,j_3) \stackrel{b\rightarrow\infty}{\sim} \frac{b^3\alpha'^2}{(4\pi)^2\sqrt{-\mathcal{N}_0}}\left(\frac{1}{2\pi\mu}\right)^{\frac{1}{b^2}}\frac{\Gamma(\frac{1}{4}+\frac{i(\pm p_1 \pm p_2 \pm p_3)}{2})}{|\Gamma(2ip_1 )\Gamma(2ip_2)\Gamma(2ip_3)|}\bigg[1 - \frac{\gamma_{\text{E}}}{b^2} + \mathcal{O}\left(\frac{1}{b^4}\right)\bigg],
\end{equation}
where the normalization constant $\sqrt{-\mathcal{N}_0}$ is given in \eqref{normalization constant}. Obtaining explicit string star backgrounds then amounts to using this expression in \eqref{string star equation 1} and finding numerical solutions for small $\epsilon$ and large $k$. Note that we have written the subleading $\frac{1}{k-2}$ term in the three-point coefficient \eqref{3-point normalized semiclassical} because it is so simple, but it is important to remember that there are more contributions which enter at the same subleading $\frac{1}{k-2}$ order from the other cubic terms in the beta function \eqref{winding beta function general}.

Solving nonlinear integral equations is not an easy task, even numerically. To help simplify the equations, we shall make use of the fact that there exists a special temperature at which the weights $\Delta_p^{\pm}$ and $\Delta_p^{(0)}$ of the winding and graviton operators are equal. In fact, this special temperature is nothing other the \emph{flat space} Hagedorn temperature, at which $\beta_* = 4\pi\sqrt{\alpha'}$ and hence
\begin{equation}
\epsilon_* = \sqrt{\frac{k-2}{k-\frac{9}{4}}} - 1.
\end{equation}
At this temperature, the beta functions $\beta^{\pm}(p)$ and $\beta^{(0)}(p)$ in \eqref{beta winding} and \eqref{beta graviton} are identical in form, at least to this order in perturbation theory. Thus, exactly at $\epsilon_*$, the transverse graviton backreaction profile is trivially given by\footnote{One overall sign is undetermined but also unphysical, so without loss of generality we may choose $f^{(0)}_*(p) = -f^{\pm}_*(p)$ instead of $f^{(0)}_*(p) = f^{\pm}_*(p)$. This is the choice that makes $f_*^{\pm}(p)$ positive.}
\begin{equation}
f^{(0)}_*(p) = -f^{\pm}_*(p),
\end{equation}
again as long as the massive string states can be neglected, which is valid for large enough $k$. Here, $f^{\pm}_*(p)$ is the solution to the \emph{quadratic} integral equation
\begin{equation}
0 \stackrel{!}{=} (2-\Delta_p^{\pm})f^{\pm}_*(p) - 8\pi\alpha'Q^2\int_0^{\infty}\frac{dp'}{2\pi}\int_0^{\infty}\frac{dp''}{2\pi}\hat{\mathcal{C}}^{(0)}(p,p',p'')f^{\pm}_*(p')f^{\pm}_*(p'').
\end{equation}
As such, instead of constructing the string star as a double expansion in large $k$ and small $\epsilon$, it is simpler to construct the string star instead as a double expansion in large $k$ and small $\epsilon-\epsilon_*$; these two expansions are obviously identical as $k\rightarrow\infty$, but the latter provides better control at finite $k$. Explicitly, the leading string star profile $f_*(p)$ exactly at $\beta = \beta_*$ is the solution to the equation
\begin{multline}\label{quadratic equation}
0 \stackrel{!}{=} -\left(p^2 + \frac{1}{4}\right)f_*(p) 
\\ + \frac{(k-2)^{3/4}}{4\pi^2\sqrt{2}}\left(1-\frac{\gamma_{\text{E}}}{k-2}\right)\int_0^{\infty}\frac{dp'}{2\pi}\int_0^{\infty}\frac{dp''}{2\pi}\frac{\Gamma(\frac{1}{4}+\frac{i(\pm p\pm p'\pm p'')}{2})}{|\Gamma(2ip)\Gamma(2ip')\Gamma(2ip'')|}f_*(p')f_*(p'').
\end{multline}
On general grounds, we see that the amplitude of the winding profile should scale like $\frac{1}{(k-2)^{3/4}}$ and that, for a normalizable solution, $f_*(p)$ must vanish linearly as $p\rightarrow 0$ and must decay exponentially as $p\rightarrow\infty$. We shall solve \eqref{quadratic equation} numerically in the following section, where these expectations are indeed met.

Once the special solution $f^{\pm}_*(p) = -f^{(0)}_*(p) = f_*(p)$ has been found for a given $k$, the string star at other nearby temperatures is obtained by substituting
\begin{align}
f^{\pm}(p) & = f_*(p) + (\epsilon-\epsilon_*)h^{\pm}(p) + \dotsc 
\\ f^{(0)}(p) & = -f_*(p) + (\epsilon-\epsilon_*)h^{(0)}(p) + \dotsc
\end{align}
into \eqref{beta winding} and \eqref{beta graviton}, then solving for the linearized perturbations $h^{\pm}(p)$ and $h^{(0)}(p)$, and so on until exhaustion sets in. Since the only temperature dependence is in $\Delta_p^{\pm}$ and the constant prefactor of $\hat{\mathcal{C}}^{(0)}(p,p',p'')$, the form of this temperature expansion is quite simple. At first order, choosing the winding profile to be real, the winding and graviton corrections are found by solving the inhomogeneous linear integral equations
\begin{align}
\notag 0 & \stackrel{!}{=} -\left(p^2 + \frac{1}{4}\right)h^{\pm}(p) + \frac{3}{2}\sqrt{\frac{k-\frac{9}{4}}{k-2}}\left(p^2 + \frac{1}{4} - \frac{4}{3}(k-2)\right)f_*(p)
\\ \label{h equation 1} & \hspace{12pt} -\frac{(k\hspace{-2pt}-\hspace{-2pt}2)^{3/4}}{4\pi^2\sqrt{2}}\hspace{-2pt}\left(\hspace{-2pt}1\hspace{-2pt}-\hspace{-2pt}\frac{\gamma_{\text{E}}}{k\hspace{-2pt}-\hspace{-2pt}2}\hspace{-2pt}\right)\hspace{-4pt}\int_0^{\infty}\hspace{-2pt}\frac{dp'}{2\pi}\hspace{-2pt}\int_0^{\infty}\hspace{-2pt}\frac{dp''}{2\pi}\frac{\Gamma(\frac{1}{4}+\frac{i(\pm p\pm p'\pm p'')}{2})}{|\Gamma(2ip)\Gamma(2ip')\Gamma(2ip'')|}\hspace{-2pt}\left[h^{(0)}(p')-h^{\pm}(p')\right]\hspace{-2pt}f_*(p'')
\\ \notag 0 & \stackrel{!}{=} -\left(p^2 + \frac{1}{4}\right)h^{(0)}(p) - \frac{3}{2}\sqrt{\frac{k-\frac{9}{4}}{k-2}}\left(p^2 + \frac{1}{4}\right)f_*(p)
\\ \label{h equation 2} & \hspace{12pt} - \frac{(k\hspace{-2pt}-\hspace{-2pt}2)^{3/4}}{2\pi^2\sqrt{2}}\hspace{-2pt}\left(\hspace{-2pt}1\hspace{-2pt}-\hspace{-2pt}\frac{\gamma_{\text{E}}}{k\hspace{-2pt}-\hspace{-2pt}2}\hspace{-2pt}\right)\hspace{-2pt}\int_0^{\infty}\frac{dp'}{2\pi}\int_0^{\infty}\frac{dp''}{2\pi}\frac{\Gamma(\frac{1}{4}+\frac{i(\pm p\pm p'\pm p'')}{2})}{|\Gamma(2ip)\Gamma(2ip')\Gamma(2ip'')|}h^{\pm}(p')f_*(p'').
\end{align}
Proceeding iteratively, each order in the temperature expansion is described by a system of linear integral equations whose sources and integration kernels are constructed out of the lower-order solutions.

\section{Numerical Solutions}\label{numerics}

\subsection{At the Special Temperature}

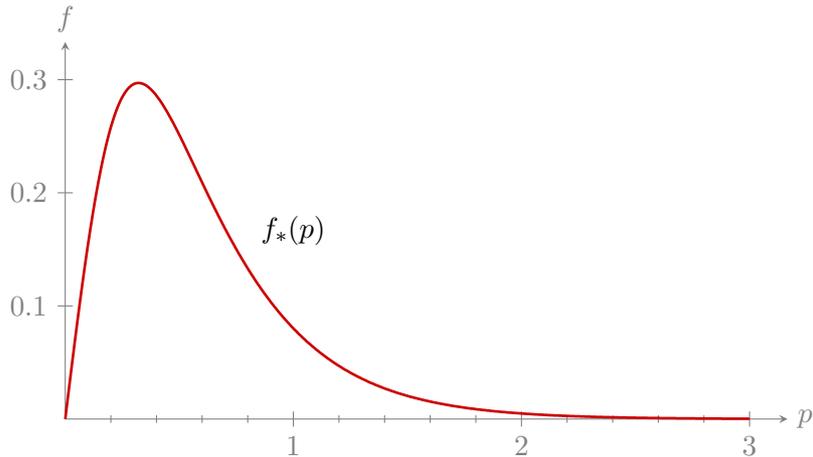
\begin{figure}
\centering
\begin{tikzpicture}
\draw[gray,-stealth] (0,0) -- (9.5,0) node[right]{$p$};
\draw[gray,-stealth] (0,0) -- (0,5) node[above]{$f$};
\draw[gray] (3,0.1) -- (3,-0.1) node[below]{$1$};
\draw[gray] (6,0.1) -- (6,-0.1) node[below]{$2$};
\draw[gray] (9,0.1) -- (9,-0.1) node[below]{$3$};
\foreach \x in {1,2,3,4,6,7,8,9,11,12,13,14}{
\draw[gray] ({0.6*\x},0.05) -- ({0.6*\x},-0.05);
}
\draw[gray] (0.1,1.5) -- (-0.1,1.5) node[left]{$0.1$};
\draw[gray] (0.1,3) -- (-0.1,3) node[left]{$0.2$};
\draw[gray] (0.1,4.5) -- (-0.1,4.5) node[left]{$0.3$};
\draw[black!20!red,line width = 1pt] plot[smooth,yscale={150*0.98696},xscale=3] file {AdS_gp_k20.dat};
\node (f) at (3,2.5) {$f_*(p)$};
\end{tikzpicture}
\caption{The string star winding and graviton profile $f_*(p)$ for $k=20$ at the flat space Hagedorn inverse temperature $\beta_* = 4\pi\sqrt{\alpha'}$, which numerically solves the integral equation \eqref{quadratic equation}.}
\label{gp profile}
\end{figure}

In the semi-classical limit, the equation \eqref{quadratic equation} for the special profile $f_*(p)$ at the flat space Hagedorn temperature depends on $k$ only through the overall coefficient of the integral term, and hence $f_*(p)$ itself only depends on $k$ by the multiplicative constant $\frac{1}{(k-2)^{3/4}}$. Such behavior is in fact necessary because in the semiclassical limit $k\rightarrow\infty$, $k$ is no longer a tunable parameter. Thus, the leading string star profile in the semiclassical limit follows from solving \eqref{quadratic equation} numerically for any one value of $k$. We arbitrarily choose $k = 20$, and the numerical solution $f_*(p)$ is plotted in Figure \ref{gp profile}.
The norm of the semiclassical $f_*(p)$ is then found to be
\begin{equation}
\int_0^{\infty}dp \ f_*(p)^2 \approx \frac{3.353}{(k-2)^{3/2}};
\end{equation}
here, we do not write the factor of $1 - \frac{\gamma_{\text{E}}}{k-2}$ since it is not the only contribution to the equations at subleading order $\frac{1}{k-2}$.

\subsection{At Nearby Temperatures}

\begin{figure}
\centering
\begin{tikzpicture}
\draw[gray,-stealth] (0,0) -- (9.5,0) node[right]{$p$};
\draw[gray,stealth-stealth] (0,-5) -- (0,3.5) node[above]{$h$};
\draw[gray] (3,0.1) -- (3,-0.1) node[below]{$1$};
\draw[gray] (6,0.1) -- (6,-0.1) node[below]{$2$};
\draw[gray] (9,0.1) -- (9,-0.1) node[below]{$3$};
\foreach \x in {1,2,3,4,6,7,8,9,11,12,13,14}{
\draw[gray] ({0.6*\x},0.05) -- ({0.6*\x},-0.05);
}
\draw[gray] (0.1,1.5) -- (-0.1,1.5) node[left]{$10$};
\draw[gray] (0.1,3) -- (-0.1,3) node[left]{$20$};
\draw[gray] (0.1,-1.5) -- (-0.1,-1.5) node[left]{$-10$};
\draw[gray] (0.1,-3) -- (-0.1,-3) node[left]{$-20$};
\draw[gray] (0.1,-4.5) -- (-0.1,-4.5) node[left]{$-30$};
\draw[blue!10!green,line width=1pt] plot[smooth,yscale={1.5*0.98696},xscale=3] file {AdS_hw_k5.dat};
\draw[blue!10!green,line width=1pt] plot[smooth,yscale={1.5*0.98696},xscale=3] file {AdS_h0_k5.dat};
\draw[blue!20!green,line width=1pt] plot[smooth,yscale={1.5*0.98696},xscale=3] file {AdS_hw_k10.dat};
\draw[blue!20!green,line width=1pt] plot[smooth,yscale={1.5*0.98696},xscale=3] file {AdS_h0_k10.dat};
\draw[blue!30!green,line width=1pt] plot[smooth,yscale={1.5*0.98696},xscale=3] file {AdS_hw_k20.dat};
\draw[blue!30!green,line width=1pt] plot[smooth,yscale={1.5*0.98696},xscale=3] file {AdS_h0_k20.dat};
\draw[blue!40!green,line width=1pt] plot[smooth,yscale={1.5*0.98696},xscale=3] file {AdS_hw_k40.dat};
\draw[blue!40!green,line width=1pt] plot[smooth,yscale={1.5*0.98696},xscale=3] file {AdS_h0_k40.dat};
\draw[blue!50!green,line width=1pt] plot[smooth,yscale={1.5*0.98696},xscale=3] file {AdS_hw_k100.dat};
\draw[blue!50!green,line width=1pt] plot[smooth,yscale={1.5*0.98696},xscale=3] file {AdS_h0_k100.dat};
\node[scale=1] (hw) at (3.5,2.25) {$h^{\pm}(p)$};
\node[scale=1] (h0) at (3.5,-2.25) {$h^{(0)}(p)$};
\end{tikzpicture}
\caption{``The Garlic Bulb.'' The linearized temperature perturbations $h^{\pm}(p)$ and $h^{(0)}(p)$ for different values of $k$. Proceeding from smaller amplitude to larger amplitude, the values plotted here are $k = 5,10,20,40,100$.}
\label{h profiles}
\end{figure}
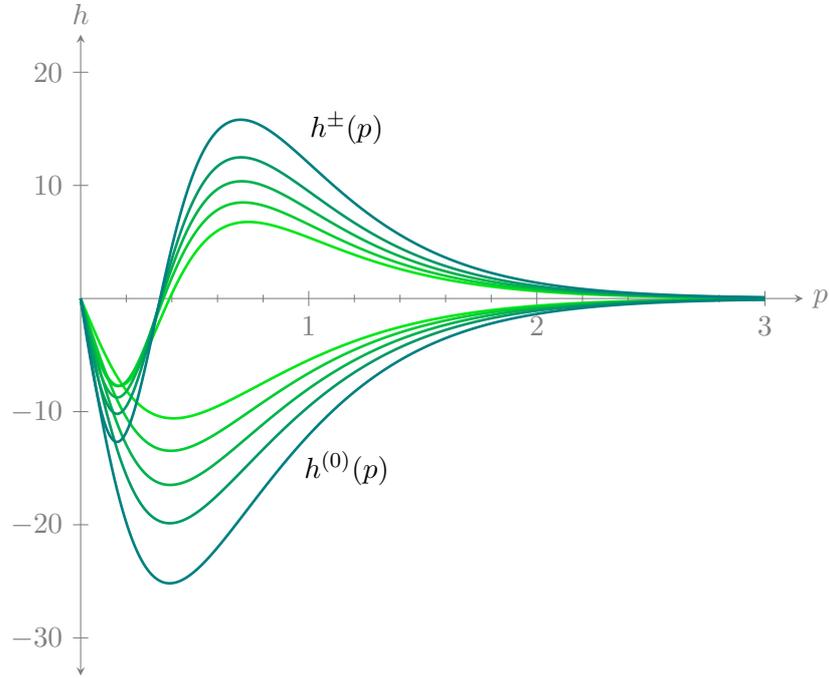

\begin{figure}
\centering
\begin{tikzpicture}
\draw[gray,-stealth] (0,0) -- (8,0) node[right]{$k$};
\draw[gray,-stealth] (0,0) -- (0,5) node[above]{$\mathrm{max}\s\s|h|$};
\draw[gray] (1.5,0.1) -- (1.5,-0.1) node[below]{$20$};
\draw[gray] (3,0.1) -- (3,-0.1) node[below]{$40$};
\draw[gray] (4.5,0.1) -- (4.5,-0.1) node[below]{$60$};
\draw[gray] (6,0.1) -- (6,-0.1) node[below]{$80$};
\draw[gray] (7.5,0.1) -- (7.5,-0.1) node[below]{$100$};
\foreach \x in {1,2,3,5,6,7,9,10,11,13,14,15,17,18,19}{
\draw[gray] ({(1.5/4)*\x},0.05) -- ({(1.5/4)*\x},-0.05);
}
\draw[gray] (0.1,1.5) -- (-0.1,1.5) node[left]{$10$};
\draw[gray] (0.1,3) -- (-0.1,3) node[left]{$20$};
\draw[gray] (0.1,4.5) -- (-0.1,4.5) node[left]{$30$};
\filldraw[red] ({1.5/4},{1.5*0.6857*0.98696}) circle (0.04);
\filldraw[red] ({1.5/2},{1.5*0.8603*0.98696}) circle (0.04);
\filldraw[red] ({1.5},{1.5*1.0502*0.98696}) circle (0.04);
\filldraw[red] ({1.5*1.5},{1.5*1.1720*0.98696}) circle (0.04);
\filldraw[red] ({1.5*2},{1.5*1.2645*0.98696}) circle (0.04);
\filldraw[red] ({1.5*2.5},{1.5*1.3404*0.98696}) circle (0.04);
\filldraw[red] ({1.5*3},{1.5*1.4051*0.98696}) circle (0.04);
\filldraw[red] ({1.5*3.5},{1.5*1.4620*0.98696}) circle (0.04);
\filldraw[red] ({1.5*4},{1.5*1.5130*0.98696}) circle (0.04);
\filldraw[red] ({1.5*4.5},{1.5*1.5592*0.98696}) circle (0.04);
\filldraw[red] ({1.5*5},{1.5*1.6017*0.98696}) circle (0.04);
\filldraw[blue] ({1.5/4},{1.5*1.0735*0.98696}) circle (0.04);
\filldraw[blue] ({1.5/2},{1.5*1.3640*0.98696}) circle (0.04);
\filldraw[blue] ({1.5},{1.5*1.6698*0.98696}) circle (0.04);
\filldraw[blue] ({1.5*1.5},{1.5*1.8648*0.98696}) circle (0.04);
\filldraw[blue] ({1.5*2},{1.5*2.0128*0.98696}) circle (0.04);
\filldraw[blue] ({1.5*2.5},{1.5*2.1339*0.98696}) circle (0.04);
\filldraw[blue] ({1.5*3},{1.5*2.2373*0.98696}) circle (0.04);
\filldraw[blue] ({1.5*3.5},{1.5*2.3281*0.98696}) circle (0.04);
\filldraw[blue] ({1.5*4},{1.5*2.4094*0.98696}) circle (0.04);
\filldraw[blue] ({1.5*4.5},{1.5*2.4831*0.98696}) circle (0.04);
\filldraw[blue] ({1.5*5},{1.5*2.5509*0.98696}) circle (0.04);
\draw[red,line width=0.7pt,variable=\k,domain=2:10] plot[smooth,samples=100] ({\k*7.5/105},{1.5*0.503*0.98696*pow(\k-2,0.25)});
\draw[red,line width=0.7pt,variable=\k,domain=10:110] plot[smooth,samples=100] ({\k*7.5/105},{1.5*0.503*0.98696*pow(\k-2,0.25)}) node[right,scale=0.9]{$5.03(k-2)^{1/4}$};
\draw[blue,line width=0.7pt,variable=\k,domain=2:10] plot[smooth,samples=100] ({\k*7.5/105},{1.5*0.800*0.98696*pow(\k-2,0.25)});
\draw[blue,line width=0.7pt,variable=\k,domain=10:110] plot[smooth,samples=100] ({\k*7.5/105},{1.5*0.800*0.98696*pow(\k-2,0.25)}) node[right,scale=0.9]{$8.00(k-2)^{1/4}$};
\node[blue,scale=1] (h0) at (4,4) {$\mathrm{max}\s\s|h^{(0)}(p)|$};
\node[red,scale=1] (hw) at (4,1.5) {$\mathrm{max}\s\s|h^{\pm}(p)|$};
\end{tikzpicture}
\caption{The maximum values of $|h^{\pm}(p)|$ and $|h^{(0)}(p)|$ from Figure \ref{h profiles} as a function of $k$, as well as the best power law fits.}
\label{h amplitudes}
\end{figure}
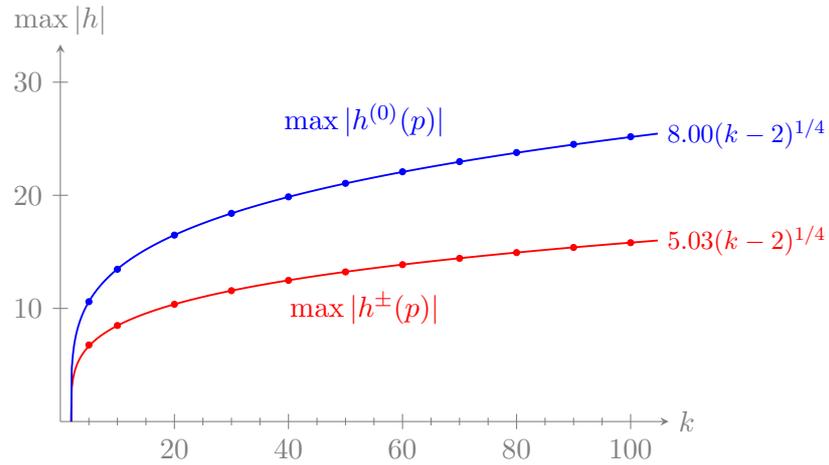

For a fixed large $k$, the string star at other temperatures near $\beta_* = 4\pi\sqrt{\alpha'}$ is found by solving \eqref{h equation 1} and \eqref{h equation 2} for the linearized perturbations $h^{\pm}(p)$ and $h^{(0)}(p)$; the winding and graviton profiles are then given to this order by $f^{\pm}(p) \approx f_*(p) + (\epsilon-\epsilon_*)h^{\pm}(p)$ and $f^{(0)}(p) \approx -f_*(p) + (\epsilon-\epsilon_*)h^{(0)}(p)$. In Figure \ref{h profiles}, we plot the numerical solutions for $h^{\pm}(p)$ and $h^{(0)}(p)$ for a range of $k$ values.
There is mild $k$-dependence in $h^{\pm}(p)$ and $h^{(0)}(p)$, primarily in their amplitudes, as before. In fact, the maximum amplitudes of $h^{\pm}(p)$ and $h^{(0)}(p)$ obey a simple power law, as shown in Figure \ref{h amplitudes}, given by $\mathrm{max}\s\s|h(p)| \stackrel{\sim}{\propto} (k-2)^{1/4}$.
Consequently, given that $f_*(p) \sim \frac{1}{(k-2)^{3/4}}$, we find that the range of temperatures over which the string star profile does not differ radically from that of $f_*(p)$ is simply
\begin{equation}
|\epsilon-\epsilon_*| \lesssim \mathcal{O}\left(\frac{1}{k-2}\right).
\end{equation}
This region is roughly equally centered around the flat space Hagedorn temperature, since $\epsilon_* \stackrel{k\rightarrow\infty}{\longrightarrow} \frac{1}{8(k-2)}$. Thus, at least in the r\'{e}gime of validity of the semiclassical limit, we have constructed reliable string star solutions in the temperature window
\begin{equation}\label{very near Hagedorn}
0 \leqslant \frac{\beta-\beta_{\text{H}}}{\beta_{\text{H}}} \lesssim \mathcal{O}\left(\frac{1}{k-2}\right),
\end{equation}
which we might deem ``very near Hagedorn'', as opposed to ``near Hagedorn'' typically referring to $\frac{\beta-\beta_{\text{H}}}{\beta_{\text{H}}} \ll 1$. However, it should be emphasized that \eqref{very near Hagedorn} is a conservative lower bound, as it simply indicates approximately where our na\"{i}ve expectations for the amplitudes of the string star component profiles, as constructed from perturbing the flat space Hagedorn temperature solution, break down. At this point, one should solve the original cubic nonlinear integral equation \eqref{string star equation 1}, which is applicable to the more broadly defined ``near Hagedorn'' region.

\begin{figure}
\centering
\begin{tikzpicture}
\draw[gray,-stealth] (0,0) -- (9.5,0) node[right]{$p$};
\draw[gray,stealth-stealth] (0,-3.5) -- (0,5) node[above]{$f$};
\draw[gray] (3,0.1) -- (3,-0.1) node[below]{$1$};
\draw[gray] (6,0.1) -- (6,-0.1) node[below]{$2$};
\draw[gray] (9,0.1) -- (9,-0.1) node[below]{$3$};
\foreach \x in {1,2,3,4,6,7,8,9,11,12,13,14}{
\draw[gray] ({0.6*\x},0.05) -- ({0.6*\x},-0.05);
}
\draw[gray] (0.1,1.5) -- (-0.1,1.5) node[left]{$0.2$};
\draw[gray] (0.1,3) -- (-0.1,3) node[left]{$0.4$};
\draw[gray] (0.1,4.5) -- (-0.1,4.5) node[left]{$0.6$};
\draw[gray] (0.1,-1.5) -- (-0.1,-1.5) node[left]{$-0.2$};
\draw[gray] (0.1,-3) -- (-0.1,-3) node[left]{$-0.4$};
\draw[red!80!green,line width=1pt] plot[smooth,yscale={75*0.98696},xscale=3] file {AdS_fw_H_k10.dat};
\draw[red!80!green,line width=1pt] plot[smooth,yscale={75*0.98696},xscale=3] file {AdS_f0_H_k10.dat};
\draw[red!60!green,line width=1pt] plot[smooth,yscale={75*0.98696},xscale=3] file {AdS_fw_H_k20.dat};
\draw[red!60!green,line width=1pt] plot[smooth,yscale={75*0.98696},xscale=3] file {AdS_f0_H_k20.dat};
\draw[red!40!green,line width=1pt] plot[smooth,yscale={75*0.98696},xscale=3] file {AdS_fw_H_k40.dat};
\draw[red!40!green,line width=1pt] plot[smooth,yscale={75*0.98696},xscale=3] file {AdS_f0_H_k40.dat};
\draw[red!20!green,line width=1pt] plot[smooth,yscale={75*0.98696},xscale=3] file {AdS_fw_H_k100.dat};
\draw[red!20!green,line width=1pt] plot[smooth,yscale={75*0.98696},xscale=3] file {AdS_f0_H_k100.dat};
\node[scale=1] (fw) at (2.5,2) {$f^{\pm}(p)$};
\node[scale=1] (f0) at (2.5,-2) {$f^{(0)}(p)$};
\end{tikzpicture}
\caption{The winding and graviton profiles, $f^{\pm}(p)$ and $f^{(0)}(p)$, exactly at the AdS Hagedorn inverse temperature $\beta = \beta_{\text{H}}$ for different values of $k$. Proceeding from larger amplitude to smaller amplitude, the values plotted here are $k = 10,20,40,100$.}
\label{Hagedorn string stars}
\end{figure}
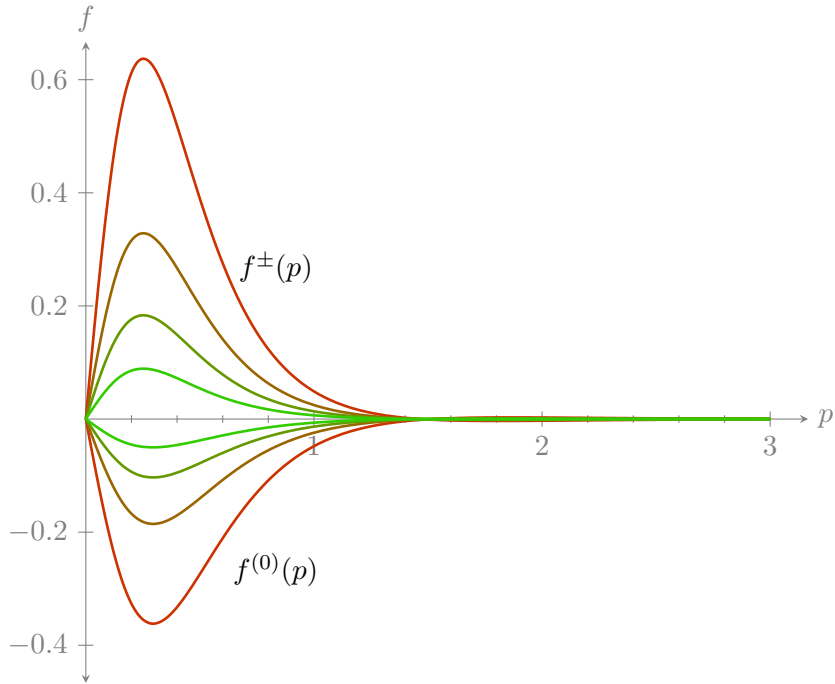

It is important to note that finite $\mathrm{AdS_3}$ string stars still exist exactly at the AdS Hagedorn temperature. These Hagedorn string star profiles are plotted in Figure \ref{Hagedorn string stars} for several different values of $k$.
As such, the $\mathrm{AdS}_3$ thermal string theory branch and the $\mathrm{AdS}_3$ string star branch as depicted in Figure \ref{string star} do not actually intersect at the Hagedorn temperature; as an immediate consequence, the other end of the correspondence line of putative string saddles also does not intersect the BTZ branch at inverse temperature $\frac{4\pi^2}{\beta_{\text{H}}}$. Moreover, there is nothing preventing us from considering the string star solution above the Hagedorn temperature as well; our backgrounds are reliable for a region for at least a region of order $\frac{1}{k-2}$ above the Hagedorn temperature. However, the stability of the above-Hagedorn string star is a more subtle question, which would require studying the perturbation theory around the new vacuum. 

\subsection{In Position Space}

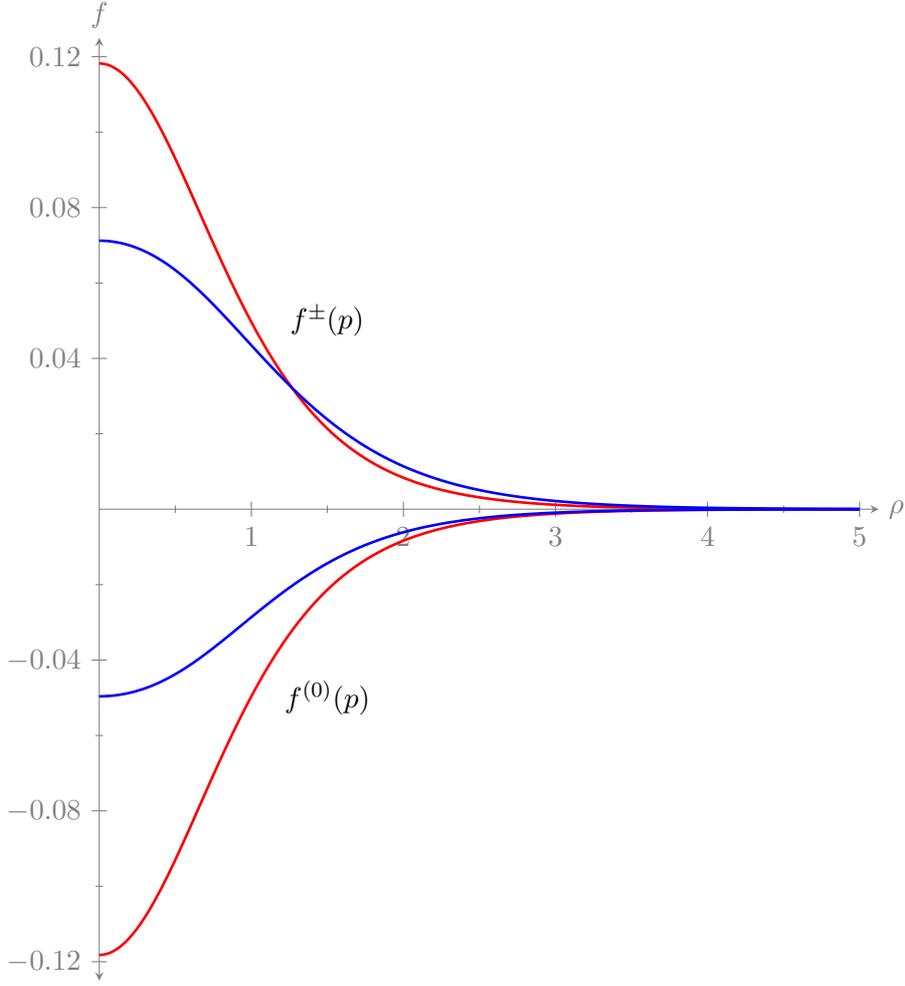
\begin{figure}
\centering
\begin{tikzpicture}
\draw[gray,-stealth] (0,0) -- (10.25,0) node[right]{$\rho$};
\draw[gray,stealth-stealth] (0,-6.25) -- (0,6.25) node[above]{$f$};
\draw[gray] (2,0.1) -- (2,-0.1) node[below]{$1$};
\draw[gray] (4,0.1) -- (4,-0.1) node[below]{$2$};
\draw[gray] (6,0.1) -- (6,-0.1) node[below]{$3$};
\draw[gray] (8,0.1) -- (8,-0.1) node[below]{$4$};
\draw[gray] (10,0.1) -- (10,-0.1) node[below]{$5$};
\foreach \x in {1,3,5,7,9}{
\draw[gray] ({\x},0.05) -- ({\x},-0.05);
}
\draw[gray] (0.1,6) -- (-0.1,6) node[left]{$0.12$};
\draw[gray] (0.1,4) -- (-0.1,4) node[left]{$0.08$};
\draw[gray] (0.1,2) -- (-0.1,2) node[left]{$0.04$};
\draw[gray] (0.1,-2) -- (-0.1,-2) node[left]{$-0.04$};
\draw[gray] (0.1,-4) -- (-0.1,-4) node[left]{$-0.08$};
\draw[gray] (0.1,-6) -- (-0.1,-6) node[left]{$-0.12$};
\foreach \x in {-5,-3,-1,1,3,5}{
\draw[gray] (0.05,{\x}) -- (-0.05,{\x});
}
\draw[red,line width=1pt] plot[smooth,yscale=50,xscale=2] file {AdS_gx_k20.dat};
\draw[red,line width=1pt] plot[smooth,yscale=-50,xscale=2] file {AdS_gx_k20.dat};
\draw[blue,line width=1pt] plot[smooth,yscale=50,xscale=2] file {AdS_fxw_H_k20.dat};
\draw[blue,line width=1pt] plot[smooth,yscale=50,xscale=2] file {AdS_fx0_H_k20.dat};
\node[scale=1] (fw) at (3,2.5) {$f^{\pm}(p)$};
\node[scale=1] (f0) at (3,-2.5) {$f^{(0)}(p)$};
\end{tikzpicture}
\caption{The $k=20$ string star profiles as functions of the $\mathrm{AdS_3}$ dimensionless radius at the flat space Hagedorn inverse temperature $\beta = 4\pi\sqrt{\alpha'}$ (red) and at the $\mathrm{AdS_3}$ Hagedorn inverse temperature $\beta = 4\pi\sqrt{\alpha'(1-\frac{1}{4(k-2)})}$ (blue).}
\label{position space}
\end{figure}

Finally, it is instructive to plot the string star profiles in position space as well. Whereas momentum space is the most natural arena for constructing Horowitz-Polchinski backgrounds from the worldsheet perspective, position space is the most natural arena from the target space effective action perspective. The momentum space and position space approaches are essentially just related by a change of operator basis, from a worldsheet energy eigenbasis (momentum space) to a basis of operators which look local in target spacetime (position basis). As familiar from quantum mechanics, these bases are related by an integral transform whose kernel is a solution to the AdS-Laplace equation in the appropriate variables. The global metric on Euclidean $\mathrm{AdS}_3$ in dimensionless coordinates is $ds^2 = \ell_{\mathrm{AdS}}^2(\cosh^2\rho\s\s d\tau^2 + d\rho^2 + \sinh^2\rho \s\s d\phi^2)$. The full AdS-Laplace operator is
\begin{equation}
\nabla^2_{\text{AdS}_3} = \frac{1}{\ell_{\text{AdS}}^2\cosh^2\rho}\frac{\partial^2}{\partial\tau^2} + \frac{1}{\ell_{\text{AdS}}^2\sinh\rho\cosh\rho}\frac{\partial}{\partial\rho}\left(\sinh\rho\cosh\rho\frac{\partial}{\partial\rho}\right) + \frac{1}{\ell_{\text{AdS}}^2\sinh^2\rho}\frac{\partial^2}{\partial\phi^2},
\end{equation}
whose zero-frequency eigenstates we seek. The complete basis of spatial functions labeled by (dimensionless) radial momentum $p$ and angular momentum $m$ is spanned by $\varphi_{p,m}(\rho)e^{im\phi}$, where the radial function satisfies
\begin{equation}
\frac{1}{\sinh\rho\cosh\rho}\frac{d}{d\rho}\left(\sinh\rho\cosh\rho\frac{d\varphi_{p,m}(\rho)}{d\rho}\right) - \frac{m^2}{\sinh^2\rho}\varphi_{p,m}(\rho) = -\left(1 + 4p^2\right)\varphi_{p,m}(\rho).
\end{equation}
The solution for the angularly symmetric (i.e.~$m=0$) case is
\begin{equation}
\varphi_{p}(\rho) = \frac{2\sqrt{2\pi}}{\cosh^{1+2ip}\hspace{-2pt}\rho}\left|\frac{\Gamma(\frac{1}{2}+ip)}{\Gamma(ip)}\right|\s\s{}_2F_1\hspace{-2pt}\left(\frac{1}{2}+ip,\frac{1}{2}+ip;1\bigg|\tanh^2\rho\right)
\end{equation}
This function is real and is normalized according to
\begin{equation}
\int_0^{\infty}d\rho \ \sinh\rho \cosh\rho \ \varphi_{p}(\rho)\varphi_{p'}(\rho) = 2\pi\delta(p-p'),
\end{equation}
for $p,p' \in \mathds{R}^+$. In order to obtain the correct normalization above, the main integral identity we derived is
\begin{multline}
\int_0^{1}\frac{dx}{(1\hspace{-2pt}-\hspace{-2pt}x)^{1-i(p+p')}}{}_2F_1\hspace{-2pt}\left(\hspace{-2pt}\frac{1}{2}\hspace{-2pt}+\hspace{-2pt}ip,\frac{1}{2}\hspace{-2pt}+\hspace{-2pt}ip;1\bigg|\s\s x\right){}_2F_1\hspace{-2pt}\left(\hspace{-2pt}\frac{1}{2}\hspace{-2pt}+\hspace{-2pt}ip',\frac{1}{2}\hspace{-2pt}+\hspace{-2pt}ip';1\bigg|\s\s x\right)
\\ = \frac{\Gamma(ip)\Gamma(-ip)}{2\Gamma(\frac{1}{2}+ip)\Gamma(\frac{1}{2}-ip)}\delta(p-p'),
\end{multline}
again for $p,p' \in \mathds{R}^+$. To establish this result, it suffices to note that the orthogonality is guaranteed by the Hermiticity of the AdS-Laplace operator, and the precise function of momentum then follows from the integrals\footnote{A.P.~Prudnikov, Yu.~A.~Brychkov and O.I.~Marichev, \emph{Integrals and Series}, \emph{Volume 3 -- More Special Functions}, subsection 2.21.1, identity 6 for the latter and subsection 2.21.10, identity 10 for the former, which also requires using the Burchnall reduction formula for the Appell function.}
\begin{align}
\int_{0}^{\infty}dp\frac{\Gamma(\frac{1}{2}+ip)\Gamma(\frac{1}{2}-ip)}{\cosh^{2ip}\rho}{}_2F_1\left(\frac{1}{2}+ip,\frac{1}{2}+ip;1\bigg|\tanh^2\rho\right) & = \frac{\pi}{2}
\\ \int_0^1 dx (1-x)^{-1+ip}{}_2F_1\left(\frac{1}{2}+ip,\frac{1}{2}+ip;1\bigg|x\right) & = \frac{\Gamma(ip)\Gamma(-ip)}{\pi}.
\end{align}
Therefore, the $\mathrm{AdS_3}$ radial momentum profiles $f(p)$ are related to the global $\mathrm{AdS_3}$ spatial profiles $\tilde{f}(\rho)$ via
\begin{align}\label{AdS position basis}
\tilde{f}(\rho) & = \int_0^{\infty}\frac{dp}{2\pi}f(p)\varphi_{p}(\rho) = \sqrt{\frac{2}{\pi}}\int_0^{\infty}\hspace{-4pt}\frac{dp \ f(p)}{\cosh^{1+2ip}\hspace{-2pt}\rho}\left|\frac{\Gamma(\frac{1}{2}\hspace{-2pt}+\hspace{-2pt}ip)}{\Gamma(ip)}\right|{}_2F_1\hspace{-2pt}\left(\frac{1}{2}\hspace{-2pt}+\hspace{-2pt}ip,\frac{1}{2}\hspace{-2pt}+\hspace{-2pt}ip;1\bigg|\tanh^2\rho\right)
\\ f(p) & = \int_0^{\infty}d\rho \s\s \sinh\rho\cosh\rho \ \tilde{f}(\rho)\varphi_p(\rho).
\end{align}
The $\sinh\rho \cosh\rho$ measure is placed entirely in the inverse transform, as opposed to being split between the two, so that a momentum profile $f(p) = 2\pi\delta(p-p_0)$ corresponds to a position space profile $\tilde{f}(\rho)$ which has definite radial momentum $p_0$ and \emph{vice versa}. In Figure \ref{position space}, we use \eqref{AdS position basis} to plot the $\mathrm{AdS_3}$ radial position profiles of the string star for $k=20$ at the flat space Hagedorn temperature as well as at the $\mathrm{AdS_3}$ Hagedorn temperature.

\section{Stringy Corrections}\label{stringy}

As $k$ gets smaller, we enter the stringy r\'{e}gime in which the massive string states become important. Here, we compute the first several relevant vertex operators in the oscillator tower and their three-point coefficients. 

All the massive string states are constructed out of current algebra descendants of the normalized tachyon vertex operator $\mathcal{O}^{w=0}_p$ in \eqref{correct basis}. Moreover, by the symmetries of the Horowitz-Polchinski set-up, the only states which can contribute to the $\mathrm{AdS_3}$ string star are those constructed out of $J^3$ (the Cartan generator) and $T$ (the quadratic Casimir). Importantly, all such string states and their three-point functions are determined in terms of \eqref{3-point normalized} by symmetry alone. Nevertheless, it quickly becomes burdensome to perform the computations by hand.

We already know that the unique massless string state which contributes is
\begin{equation}
\mathcal{O}^{(0)}_p = \frac{2}{k}J^3\widetilde{J}^{\s 3}\mathcal{O}^{0}_p,
\end{equation}
whose three-point coefficient $\hat{\mathcal{C}}^{(0)}(p_1,p_2,p_3)$ with the winding operators we already computed in \eqref{ww0}. The three-point function of $\mathcal{O}^{(0)}_p$ with itself trivially vanishes because $J^3$ is a $\mathrm{U}(1)$ current. 

At the first massive level, there are two scalar Virasoro primaries, which we denote $\mathcal{O}^{(1,1)}_p$ and $\mathcal{O}^{(1,2)}_p$, that can be constructed as current algebra descendants of the tachyon $\mathcal{O}^0_p$ obeying the symmetries. There is a natural basis to choose, as one combination still involves only transverse oscillator modes. Defining $h_p \equiv \frac{1+4p^2}{4(k-2)}$ to be the holomorphic weight of the tachyon $\mathcal{O}^0_p$, the simpler of the two first-massive vertex operators reads
\begin{equation}
\mathcal{O}^{(1,1)}_p = \frac{1}{h_p(h_p+1)k}\Big(h_p\partial J^3 - J^3\partial\Big)\Big(h_p\barred{\partial}\widetilde{J}^{\s 3}-\widetilde{J}^{\s 3}\barred{\partial}\Big)\mathcal{O}^0_p,
\end{equation}
with the prefactor again chosen for the canonical normalization used before. The second vertex operator at the first massive level is given by
\begin{equation}
\mathcal{O}_p^{(1,2)} = \frac{\normal{[(2h_p\hspace{-3pt}+\hspace{-3pt}1)T \hspace{-3pt}+\hspace{-3pt} \frac{(2h_p\hspace{-1pt}+\hspace{-1pt}1)c+2h_p(8h_p\hspace{-1pt}-\hspace{-1pt}5)}{k}\hspace{-1pt}(\hspace{-1pt}J^3\hspace{-1pt})^2\hspace{-3pt}-\hspace{-3pt}\frac{3}{2}\partial^2][(2h_p\hspace{-3pt}+\hspace{-3pt}1)\widetilde{T} \hspace{-3pt}+\hspace{-3pt} \frac{(2h_p\hspace{-1pt}+\hspace{-1pt}1)c+2h_p(8h_p\hspace{-1pt}-\hspace{-1pt}5)}{k}\hspace{-1pt}(\hspace{-1pt}\widetilde{J}^3\hspace{-1pt})^2\hspace{-3pt}-\hspace{-3pt}\frac{3}{2}\barred{\partial}^2]\hspace{-1pt}\mathcal{O}^0_p}}{(2h_p+1)[(2h_p+1)(c-1)+4h_p(8h_p-5)]\frac{c}{2}+h_p(8h_p-5)[2h_p(8h_p-5)-(2h_p+1)]},
\end{equation}
where an expression like $\normal{(J^3)^2\mathcal{O}}$ means $\normal{J^3\normal{J^3\mathcal{O}}}$, and $c = \frac{3k}{k-2}$ is the central charge of the CFT. These first-massive vertex operators have dimension
\begin{equation}
\Delta_p^{(1)} = 4 + \frac{1+4p^2}{2(k-2)}.
\end{equation}
It is then straightforward to compute the three-point coefficients of $\mathcal{O}^w_{p_1}$ and $\mathcal{O}^{-w}_{p_2}$ with $\mathcal{O}^{(1,i)}_{p_3}$ in terms of $\mathcal{C}$ given in \eqref{3-point normalized}. Letting $h^w_p = \frac{1+4p^2}{4(k-2)} + \frac{w^2\beta^2}{16\pi^2\alpha'}$ be the holomorphic weight of a winding operator, we find these three-point coefficients to be
\begin{align}
\mathcal{C}^{(1,1)} & = \frac{w^2\beta^2(h_{p_1}^w-h_{p_2}^w)^2}{16\pi^2\alpha'h_{p_3}(h_{p_3}+1)}\mathcal{C}
\\ \mathcal{C}^{(1,2)} & = \frac{\big[h_{p_3}(h_{p_3}\hspace{-5pt}-\hspace{-2pt}1)\hspace{-2pt}+\hspace{-2pt}(h_{p_1}^w \hspace{-5pt}+\hspace{-2pt} h_{p_2}^w)(2h_{p_3}\hspace{-5pt}+\hspace{-2pt}1)\hspace{-2pt}-\hspace{-2pt}3(h_{p_1}^w \hspace{-5pt}-\hspace{-2pt} h_{p_2}^w)^2 \hspace{-2pt}-\hspace{-2pt} \frac{w^2\beta^2}{8\pi^2\alpha'}[(2h_{p_3}\hspace{-5pt}+\hspace{-2pt}1)c\hspace{-2pt}+\hspace{-2pt}2h_{p_3}\hspace{-2pt}(8h_{p_3}\hspace{-5pt}-\hspace{-2pt}5)]\big]^2}{2(2h_{p_3}\hspace{-5pt}+\hspace{-2pt}1)[(2h_{p_3}\hspace{-5pt}+\hspace{-2pt}1)(c\hspace{-2pt}-\hspace{-2pt}1)\hspace{-2pt}+\hspace{-2pt}4h_{p_3}(8h_{p_3}\hspace{-5pt}-\hspace{-2pt}5)]c\hspace{-2pt}+\hspace{-2pt}4h_{p_3}(8h_{p_3}\hspace{-5pt}-\hspace{-2pt}5)[2h_{p_3}(8h_{p_3}\hspace{-5pt}-\hspace{-2pt}5)\hspace{-2pt}-\hspace{-2pt}(2h_{p_3}\hspace{-5pt}+\hspace{-2pt}1)]}\mathcal{C}.
\end{align}
Note that again the three-point function of $\mathcal{O}^{(1,1)}_p$ with itself vanishes due to involving only the transverse $\mathrm{U}(1)$ current; however, the three-point coefficients of $\langle\mathcal{O}^{(1,1)}_{p_1}\mathcal{O}^{(1,1)}_{p_2}\mathcal{O}^{(1,2)}_{p_3}\rangle$ and of $\langle\mathcal{O}^{(1,2)}_{p_1}\mathcal{O}^{(1,2)}_{p_2}\mathcal{O}^{(1,2)}_{p_3}\rangle$ are nontrivial. Nevertheless, these latter two contributions are further subleading whenever the massive string profiles are subleading to the winding and graviton profiles.

At the second massive level, there are again two Virasoro primaries consistent with the symmetries, both of which involve longitudinal modes. Without computing the normalization constants, they take the form
\begin{align}
\mathcal{O}_p^{(2,1)} & = \mathcal{N}^{(2,1)}\normal{D^{(2,1)}\widetilde{D}^{(2,1)}\mathcal{O}_p^0}
\\ \mathcal{O}_p^{(2,2)} & = \mathcal{N}^{(2,2)}\normal{D^{(2,2)}\widetilde{D}^{(2,2)}\mathcal{O}_p^0},
\end{align}
where
\begin{align}
D^{(2,1)} & = (2h_p\hspace{-2pt}+\hspace{-2pt}1)TJ^3 \hspace{-2pt}+\hspace{-2pt} \frac{5}{6}(2h_p\hspace{-2pt}+\hspace{-2pt}1)\partial^2 J^3 \hspace{-2pt}+\hspace{-2pt} \frac{5}{2}J^3\partial^2 \hspace{-2pt}+\hspace{-2pt} \frac{(2h_p\hspace{-2pt}+\hspace{-2pt}1)c\hspace{-2pt}+\hspace{-2pt}16h_p^2\hspace{-2pt}+\hspace{-2pt}74h_p\hspace{-2pt}+\hspace{-2pt}18}{3k}(J^3)^3
\\ D^{(2,2)} & = \partial^3 \hspace{-2pt}+\hspace{-2pt} h_p(h_p\hspace{-2pt}+\hspace{-2pt}1)\partial T \hspace{-2pt}-\hspace{-2pt} 2(h_p\hspace{-2pt}+\hspace{-2pt}1)T\partial \hspace{-2pt}+\hspace{-2pt} \frac{2[(h_p\hspace{-2pt}+\hspace{-2pt}1)c\hspace{-2pt}+\hspace{-2pt}3h_p^2\hspace{-2pt}-\hspace{-2pt}7h_p\hspace{-2pt}+\hspace{-2pt}2]}{k}[h_p \partial J^3 J^3 \hspace{-2pt}-\hspace{-2pt} (J^3)^2\partial],
\end{align}
and again right-nested normal ordering is taken throughout.

\section{Summary}

In conclusion, we have studied $\mathrm{AdS_3}$ string stars at pure NSNS flux using the Horowitz-Polchinski methodology adapted to the worldsheet. We have shown how to use the $\alpha'$-exact worldsheet CFT to construct the string star equations to any desired accuracy in a double expansion in $\frac{1}{k-2}$ and $\beta-\beta_{\text{H}}$. We also provided explicit numerical solutions to these equations in the r\'{e}gime where massive string corrections can be neglected as well as the formulae necessary to obtain the first-massive stringy corrections.

It would be interesting to obtain a quantitative understanding of the stringy corrections in order to probe the very stringy region $k-3 \simeq \mathcal{O}(1)$. In this r\'{e}gime, one must include the higher-order contributions to the beta function itself. It seems that the easiest way to obtain the correct equations to one higher order in the coupling functions is to start from the full four-point function of the time-winding operators, which should automatically incorporate the lower-order contributions from the entire massive string tower. We leave this to future work.

Another avenue of future directions involves analyzing the above-Hagedorn string stars. In this paper, we demonstrated that in $\mathrm{AdS_3}$ at finite $k$ there is no obstruction in continuing the string star solutions to temperatures above the Hagedorn temperature. Two interesting open questions are whether the above-Hagedorn string stars are stable backgrounds and, if so, whether the thermal string branch flows to the string star branch above the Hagedorn temperature. For the former question, one would have to show that there are no relevant deformations that can perturb away from the string star solution. For the latter question, one would have to track the flow induced by the relevant winding string vertex operators. We hope to be able to say more about these topics in the future.

\section*{Acknowledgments}

We thank Indranil Halder for collaboration on early stages of this project. We would also like to thank Ofer Aharony, Mina Himwich and Erez Urbach for helpful conversations. This work is supported in part by DOE grant DE-SC0007870.

\appendix
\section{Properties of the Barnes Upsilon Function}\label{Barnes}

The definition of the Barnes Upsilon function $\Upsilon_b(x)$ given in the text in \eqref{Upsilon definition} was
\begin{equation}
\ln\Upsilon_b(x) = \int_0^{\infty}\frac{dt}{t}\left[\bigg(\frac{b+\frac{1}{b}}{2}-x\bigg)^2 e^{-t} - \frac{\sinh^2\hspace{-2pt}\big((\frac{b+\frac{1}{b}}{2}-x)\frac{t}{2}\big)}{\sinh(\frac{bt}{2})\sinh(\frac{t}{2b})}\right]
\end{equation}
for $0 < \mathrm{Re} \s\s x < b + \frac{1}{b}$, and elsewhere by analytic continuation. The Barnes Upsilon function has zeros at $x = -mb - \frac{n}{b}$ and at $x = (m+1)b + \frac{n+1}{b}$ for all $m,n \in \mathds{Z}^+$, and it satisfies the reflection relations
\begin{align}
\Upsilon_b(x) & = \Upsilon_{1/b}(x)
\\ \Upsilon_b(x) & = \Upsilon_b\left(b + \frac{1}{b} - x\right).
\end{align}
Its derivative at zero argument is $\Upsilon_b'(0) = \Upsilon_b(\frac{1}{b})$. The fundamental shift relations are
\begin{align}
\Upsilon_b(x+b) & = b^{1-2bx}\gamma(bx)\Upsilon_b(x)
\\ \Upsilon_b\left(x + \frac{1}{b}\right) & = b^{\frac{2x}{b}-1}\gamma\left(\frac{x}{b}\right)\Upsilon_b(x),
\end{align}
which iterate to
\begin{equation}
\frac{\Upsilon_b\hspace{-2pt}\left(x \hspace{-2pt}+\hspace{-2pt} \frac{m}{b} \hspace{-2pt}+\hspace{-2pt} nb\right)}{\Upsilon_b(x)} = b^{\frac{m(\hspace{-1pt}m\hspace{-1pt}-\hspace{-1pt}1\hspace{-1pt})}{b^2}\hspace{-1pt}+\hspace{-1pt}\frac{2mx}{b}\hspace{-1pt}+\hspace{-1pt}n\hspace{-1pt}-\hspace{-1pt}(\hspace{-1pt}2n\hspace{-1pt}+\hspace{-1pt}1\hspace{-1pt})m\hspace{-1pt}-\hspace{-1pt}2nxb\hspace{-1pt}-\hspace{-1pt}n(\hspace{-1pt}n\hspace{-1pt}-\hspace{-1pt}1\hspace{-1pt})b^2}\prod_{i=0}^{n-1}\hspace{-2pt}\gamma\hspace{-2pt}\left(m\hspace{-2pt}+\hspace{-2pt}xb\hspace{-2pt}+\hspace{-2pt}ib^2\right)\hspace{-2pt}\prod_{j=0}^{m-1}\hspace{-2pt}\gamma\hspace{-2pt}\left(\hspace{-2pt}\frac{x}{b}\hspace{-2pt}+\hspace{-2pt}\frac{j}{b^2}\hspace{-2pt}\right)\hspace{-2pt},
\end{equation}
for any $m,n \in \mathds{Z}^+$. All asymptotic expansions of $\Upsilon_b(x)$ to any desired accuracy can be obtained from the formulas given in \cite{Collier18}. The relevant expansions used in this paper can all be brought to the form
\begin{multline}
\ln\Upsilon_b\left(x+\frac{y}{b}\right) \stackrel{b\rightarrow \infty}{\sim} \frac{b^2}{4}\ln b - b^2\left(3\ln A - \frac{1}{12}\ln 2\right) + xb\ln b + \left(x^2 + \frac{2y-1}{2}\right)\ln b
\\ - \ln\Gamma\left(xb+y\right) + \frac{1}{2}\ln(2\pi) - x^2\gamma_{\text{E}} + \frac{x(2y-1)}{b}\ln b + \frac{(2y-1)^2}{4b^2}\ln b - \frac{x(2y-1)\gamma_{\text{E}}}{b}
\\ + \frac{1}{12b^2}\left[\gamma_{\text{E}} + \psi^{(0)}(\tfrac{1}{2}) - 3(2y-1)^2\gamma_{\text{E}} + x^4\psi^{(2)}(1)\right] + \mathcal{O}\left(\frac{1}{b^3}\right),
\end{multline}
where $\gamma_{\text{E}} = 0.577215\dotsc$ is the Euler-Mascheroni constant, $A = 1.282427\dotsc$ is Glaisher's constant and $\psi^{(n)}(z) \equiv \frac{d^{n+1}}{dz^{n+1}}\ln\Gamma(z)$ is the polygamma function.

\end{document}